\documentclass[aps,prb,reprint,showkeys]{revtex4-1}

\usepackage{graphicx}
\usepackage{hyperref}

\begin{document}

\title{Can molecular projected density-of-states (PDOS) be systematically used in electronic conductance analysis?}

\author{Tonatiuh Rangel}
\altaffiliation[Present address: ]{Molecular Foundry, Lawrence Berkeley National Laboratory, Berkeley, California 94720, USA.}
\affiliation{Institute of Condensed Matter and Nanosciences, Universit\'e catholique de Louvain, Chemin des \'Etoiles 8, bte L7.03.01, 1348 Louvain-la-Neuve, Belgium,}
\affiliation{European Theoretical Spectroscopy Facility (ETSF).}
\author{Gian-Marco Rignanese}
\affiliation{Institute of Condensed Matter and Nanosciences, Universit\'e catholique de Louvain, Chemin des \'Etoiles 8, bte L7.03.01, 1348 Louvain-la-Neuve, Belgium,}
\affiliation{European Theoretical Spectroscopy Facility (ETSF).}
\author{Valerio Olevano}
\affiliation{CNRS, Institut N\'eel, F-38042 Grenoble, France,}
\affiliation{Univ. Grenoble Alpes, F-38000 Grenoble, France,}
\affiliation{European Theoretical Spectroscopy Facility (ETSF).}

\date{\today}

\begin{abstract}
Using benzene-diamine and benzene-dithiol molecular junctions as benchmarks, we investigate the widespread analysis of the quantum transport conductance $\mathcal{G}(\epsilon)$ in terms of the projected density of states (PDOS) onto molecular orbitals~(MOs).
We first consider two different methods for identifying the relevant MOs: 1) diagonalization of the Hamiltonian of the isolated molecule, and 2) diagonalization of a submatrix of the junction Hamiltonian constructed by considering only basis elements localized on the molecule.
We find that these two methods can lead to substantially different MOs and hence PDOS.
Furthermore, within Method 1, the PDOS can differ depending on the isolated molecule chosen to represent the molecular junction (e.g. benzene-dithiol or -dithiolate); and, within Method 2, the PDOS depends on the chosen basis set.
We show that these differences can be critical when the PDOS is used to provide an physical interpretation of the conductance (especially, when it has small values as it happens typically at zero bias).
In this work, we propose a new approach trying to reconcile the two traditional methods.
Though some improvements are achieved, the main problems are still unsolved.
Our results raise more general questions and doubts on a PDOS-based analysis of the conductance.

\centerline{\includegraphics[width=0.5\columnwidth]{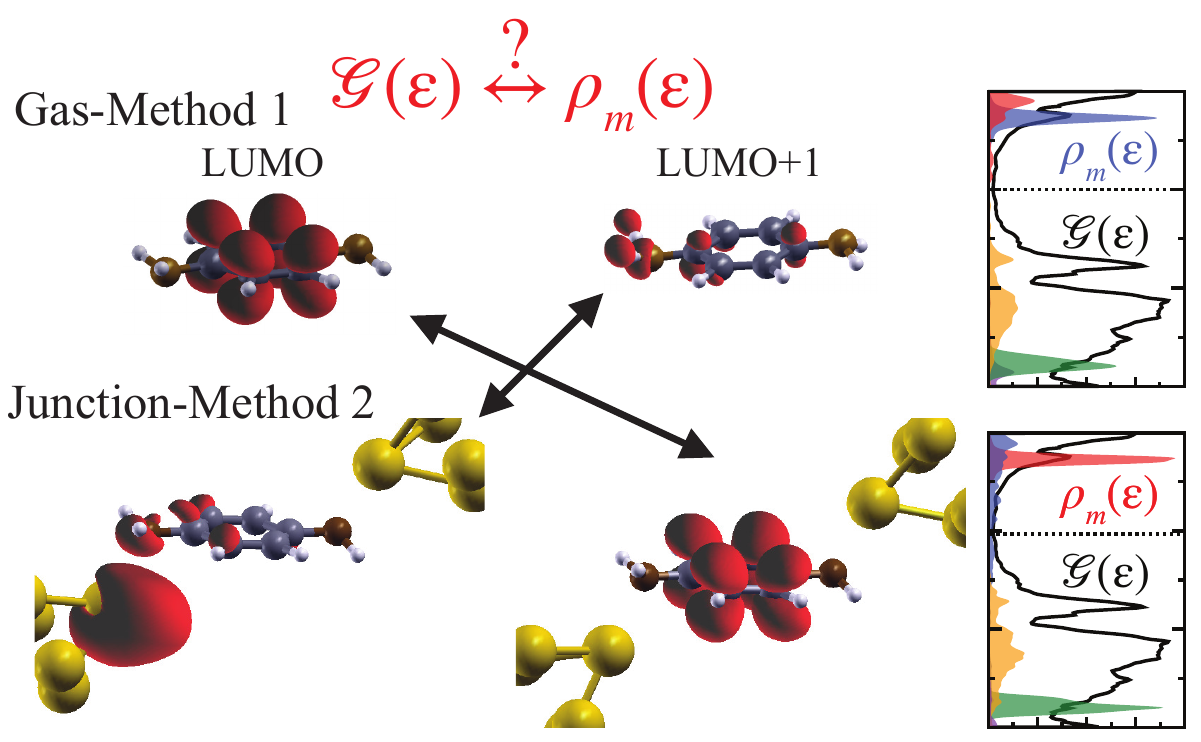}}
\end{abstract}


\maketitle

\section*{Introduction}

According to Moore's law, in a decade or so the downscaling of conventional silicon based electronics will achieve its ultimate nanoscale limits.
Molecular electronics, or electronics at the nanoscale, is considered one of the most difficult technological challenges.
The construction, measurement and understanding of electronic devices constituted by single molecules in between metal electrodes, is nowadays a major concern of fundamental research.

Today, different techniques are available to realize molecular junctions in laboratories, such as electromigration methods, mechanical strain and scanning tunneling microscopes to open small gaps in between gold leads which can host (with a small but non-negligible probability) single molecules from a wetting solution~\cite{reed_conductance_1997,xu_measurement_2003,park_nanomechanical_2000}.
The complete characterization of such junctions (including the measurement of their current-voltage characteristics) is however still difficult to achieve.
In order to obtain reliable single-molecule zero-bias conductances, it was suggested to resort to a statistically significant sample of tens of thousands measurements~\cite{venkataraman_single-molecule_2006}.
Thanks to this breakthrough work, it is now possible to quote the $0$-bias conductance of some molecular junctions such as benzene-diamine~(BDA) and benzene-dithiol~(BDT) in between gold leads.
Nevertheless, important characterization uncertainties still persist.
For instance, in these experiments the junction geometry is not measured and hence unknown.
Given these difficulties, resorting to theory could reveal a valid approach to understand and interpret the experimental observations.

The theoretical description of the electronic quantum transport in molecular junctions or nanostructures relies on established frameworks \cite{Datta,DiVentra} like the Kubo-Greenwood \cite{Kubo, Greenwood} or the Landauer \cite{Landauer} formalisms, or the non-equilibrium Green's function theory \cite{Keldysh, Danielewicz, RammerSmith}.
In the last two decades, the combination of these formalisms with density-functional theory (DFT) or many-body perturbation~(MBPT) theory allowed to establish \textit{ab initio} approaches to quantum transport.
The DFT-Landauer framework is one of the most popular.
It has proven successful in calculating zero-bias conductances in good agreement with the experiment in some systems like the hydrogen molecule in between platinum wires~\cite{ThygesenJakobsen2005}.
In other systems, like organic molecule junctions, the DFT-Landauer estimate can be several orders of magnitude larger than the experiment~\cite{NitzanRatner,reed_conductance_1997}.
Several solutions have been proposed to alleviate this discrepancy as by self-interaction corrections \cite{toher_effects_2008,pontes_ab_2011}, hybrid mixed Hartree-Fock approaches \cite{ferretti_ab_2012}, many-body model \cite{quek_aminegold_2007,darancet_quantitative_2012,dellangela_relating_2010,cehovin_role_2008} or \textit{ab initio} GW corrections \cite{strange,rangel_transport_2011}, arising in a not yet solved controversy \cite{stadler_fermi_2006, stadler_fermi_2007, stadler_conformation_2010, kastlunger_charge_2013, baldea_transition_2012, mera_are_2010, mera_assessing_2010, baldea_quantifying_2014, baldea_quantum_2014}.

Besides calculating or measuring, a physical interpretation of the conductance is needed. At the end, we would like a complete picture of the mechanisms governing quantum transport in order to fully understand the behavior of the molecular junction as an electronic device. 
To this end, it is important to establish a relationship between the conductance and the electronic structure, for example determining the main ingredients influencing the absolute value of the zero-bias conductance
A very common approach for providing such an interpretation proceeds as follows.
A set of molecular orbitals~(MOs) associated to the central molecule are identified and classified according to the energy levels, \textit{e.g.} the highest occupied molecular orbital (HOMO), or the lowest unoccupied molecular orbital (LUMO), or the next one (LUMO+1), etc.
Then, the total electronic density of states~(DOS) is decomposed into the projected density of states (PDOS) associated to each different MO.
Finally, by directly comparing the conductance profile $\mathcal{G}(\epsilon)$ with the various PDOS, one tries to establish a correspondence between conductance features and MOs.
In particular, one tries to understand which MO has the largest influence on the zero-bias conductance.

The purpose of this work is to investigate how meaningful (or on the contrary misleading) this analysis is. 
How reliable are the interpretations that one can get? How pertinent is it to a correct understanding of the behavior of the system?
We analyze two common benchmarks, the above mentioned molecular junctions of BDA and BDT in between gold leads in order to answer these questions and solve the problems evidenced in traditional methodologies.
In particular, we propose a new method to identify MOs and the associated PDOS which clearly goes in this direction, though further work is still required.
Though the findings of this work may seem quite theoretical at first sight, they will have an important impact in the experimental community. Indeed, theoretical analysis of quantum transport is often used for interpreting the measurements, predicting trends (for example, for the sign of the thermopower), for obtaining independent arguments, or checking the validity of the experimental work.

The paper is organized as follows:
Sec.~I introduces quantum transport \textit{ab initio} theory, together with the definitions of all the relevant quantities and the two traditional methods to identify MOs and PDOS.
In Sec.~II and III, we present the results for the BDA and BDT molecular junctions, respectively.
Sec.~IV  is devoted to the presentation of our new method and its results on BDT.
Sec.~V gives a critical discussion of the physical meaning of the interpretation provided by the traditional methods and our new one.

\section{Theory: Molecular orbitals, PDOS and conductance}
\label{theory}

In the DFT-Landauer framework, the molecular junction is modeled by a central region (C) connected to two semi-infinite leads (left L and right R).
Its conductance $\mathcal{G}(\epsilon)$ as a function of the energy $\epsilon$ of the injected electrons is given by the Landauer formula:
\[
  \mathcal{G}(\epsilon) = \frac{2e^2}{h} M(\epsilon)T(\epsilon)
   = \frac{2e^2}{h} \mathop{\mathrm{tr}}
   [\Gamma_L(\epsilon) G^r_C(\epsilon) \Gamma_R(\epsilon) G^a_C(\epsilon)]
   .
\]
$M(\epsilon)$ is the number of modes at a given energy $\epsilon$. $T(\epsilon)$ is their transmittance. $\Gamma_{L/R}(\epsilon)$ is the left/right leads injection rate. $G^{r/a}_C(\epsilon)$ is the retarded/advanced Green function for the central region.
The quantities $G^{r/a}_C(\epsilon)$ and $\Gamma_{L/R}(\epsilon)$ can be obtained from the DFT electronic structure [\textit{i.e.} the energies $\epsilon_n$ and wavefunctions $\phi_n(r)$] of the central region containing an ``extended molecule'' and of the leads (treated as infinite periodic solids), respectively.
The central extended molecule actually consists of the molecule itself plus some layers actually belonging to the leads.
The number of included layers (typically 3 or 4) should account for the relaxation of both the atomic and the electronic structures of the junction.
The value assumed by $\mathcal{G}(\epsilon)$ at the Fermi energy $\epsilon_F$ (which will be set to 0 in the following), $\mathcal{G}(\epsilon$=$\epsilon_F$=0), is an observable that can be directly measured in experiments and referred to as the \textit{zero-bias conductance}. 

The junction conductance depends on the nature and the shape of the leads, the geometric/atomic structure of the molecule-lead contact, and the molecule itself.
Experiments and calculations very often only consider gold for the leads, so that these can be considered a constant ingredient.
In contrast, the geometry of the molecule-lead contact may vary quite a lot, but in many cases it is not known and furthermore it cannot be controlled experimentally.
In practice, experiments only measure conductances averaged over the different possible geometries.
In the end, the main factor influencing the junction conductance is the central molecule.
Therefore, there is quite a lot of interest on how the conductance changes by varying the chemical composition or the atomic structure of the central molecule.
Furthermore, by looking at the generic representation of the molecular junction, the central molecule appears as a ``bottleneck'' to the stream of electrons flowing from one lead to the other.
For this reason, it is believed that the molecule itself and its electronic structure has a deep influence on the conductance.

The interpretation of the conductance profile $\mathcal{G}(\epsilon)$ or of the zero-bias conductance $\mathcal{G}(0)$ is often carried out by referring to the projected density-of-states onto molecular orbitals (see next Section).
Traditionally, these are identified using two methods that will be detailed later.

\subsection{Interpretation of the conductance by the PDOS}

Supposing that a set $\{m\}$ of molecular orbitals with wavefunctions $\phi^\mathrm{MO}_m(r)$ have been identified, the projected density of states, $\rho_m(\epsilon)$, on the molecular orbital $m$ is defined as
\begin{equation}
  \rho_m(\epsilon) = 
  \sum_n \langle \phi^\mathrm{MO}_m | \phi_n \rangle
  \delta(\epsilon-\epsilon_n)
  ,
  \label{PDOS}
\end{equation}
where $n$ runs over all the states of the central extended molecule with wavefunction $\phi_n(r)$ and energy $\epsilon_n$.
In Eq.~(\ref{PDOS}), the Dirac delta function is usually replaced by a Gaussian function with a given broadening.

As discussed above, $\mathcal{G}(\epsilon)$ is fundamentally determined by the electronic structure of the central extended molecule.
In particular, the DOS $\rho(\epsilon)$=$\sum_n \delta(\epsilon - \epsilon_n)$ should play a major role.
For instance, the conductance will be zero where the number of modes $M(\epsilon)$=0, and so will be the density-of-states.
Hence, it is quite natural to interpret the conductance with the help of the DOS.
More specifically, it has become very common to analyze $\mathcal{G}(\epsilon)$ in terms of the different partial \textit{molecular} components which enter the full DOS, \textit{i.e.} the PDOS on the various MOs~\cite{shi_ab_2006,gao_comparable_2013,gutierrez_conductance_2003,sen_single-molecule_2013,caliskan_spin_2014}.
Since the energy region of interest for the conductance is that one around the Fermi level, one usually takes into account the molecular orbitals around the fundamental gap, \textit{e.g.} the highest occupied molecular orbital (HOMO), the lowest unoccupied molecular orbital (LUMO), and the next ones, the LUMO+1, LUMO+2, HOMO-1, etc.

The analysis of the conductance in terms of the PDOS is based on a one-to-one comparison of $\mathcal{G}(\epsilon)$ with $\rho_m(\epsilon)$ for some chosen MOs.
Whenever a peak in $\mathcal{G}(\epsilon)$ is in correspondence with a peak in a $\rho_m(\epsilon)$, that molecular orbital $m$ is said to ``drive'' the peak of conductance.
The specific case of the $0$-bias conductance is a bit particular.
Indeed, very often, $\mathcal{G}(0)$ is quite small and the main conductance peaks are several eV away.
The $0$-bias conductance is actually interpreted as the tail of one of these peaks.
But there is some ambiguity about the MO which will be said to drive $\mathcal{G}(0)$.
Indeed, it can be chosen as:
\vspace{-\topsep}
\begin{itemize}\itemsep2pt \parskip0pt \parsep0pt
\item[i)] the MO corresponding to the peak closest to the Fermi level ($\epsilon$=0)~\cite{kaliginedi_correlations_2012,perrin_large_2014};
\item[ii)] the MO presenting the highest PDOS value at $\epsilon$=0, no matter how far the PDOS maximum is from $\epsilon$=0~\cite{rangel_transport_2011,strange_towards_2011,ning_first-principles_2007}.
\end{itemize}

\subsection{Identification of the molecular orbitals}

The molecular orbitals $\phi^\mathrm{MO}_m(r)$ are the fundamental ingredient of the PDOS [see Eq.~(\ref{PDOS})].
As shown below, the approach chosen for identifying the MOs strongly affects the PDOS and the consequent interpretation of the conductance spectrum.
Two main methods have been used so far in the literature for identifying MOs:

\begin{description}
\item[Method 1]
The $\phi^\mathrm{MO}_m(r)$ are chosen to be the eigenfunctions of Hamiltonian of the uncontacted, gas phase, isolated molecule~\cite{ning_first-principles_2007}.
For consistency, they are usually determined using exactly the same supercell of the extended molecule, as in the molecular junction calculation, and removing the atoms of the leads.

\item[Method 2]
The Hamiltonian of the extended molecule is first expressed on a real-space localized basis set.
This can be achieved, for instance, using maximally localized Wannier functions (MLWFs)~\cite{marzari_maximally_1997}.
The $\phi^\mathrm{MO}_m(r)$ are then chosen as the eigenfunctions of the submatrix constructed by considering only basis elements localized on the molecule~\cite{ThygesenJakobsen2005}.
\end{description}

\textit{There is no obvious reason why the MOs identified using these two different procedures should coincide.}
Furthermore, it is not evident which method is preferred with respect to the assumed purpose, \textit{i.e.} the analysis of the conductance.
Method 1 coincides with the rigorous definition of MOs in the chemistry sense for the isolated molecule.
However, the electronic structure of the extended molecule (taking into account charge transfer and other modifications induced by the contact between the molecule and the leads) is clearly much more important with respect to the conductance profile. 
So that Method 2 appears more relevant for the analysis of the conductance.

Note that choosing of one of these methods does not affect the conductance profile, provided that convergence is reached.
What actually changes is rather the PDOS and hence the interpretation of the conductance in these terms.

\begin{figure*}
\includegraphics[width=\textwidth]{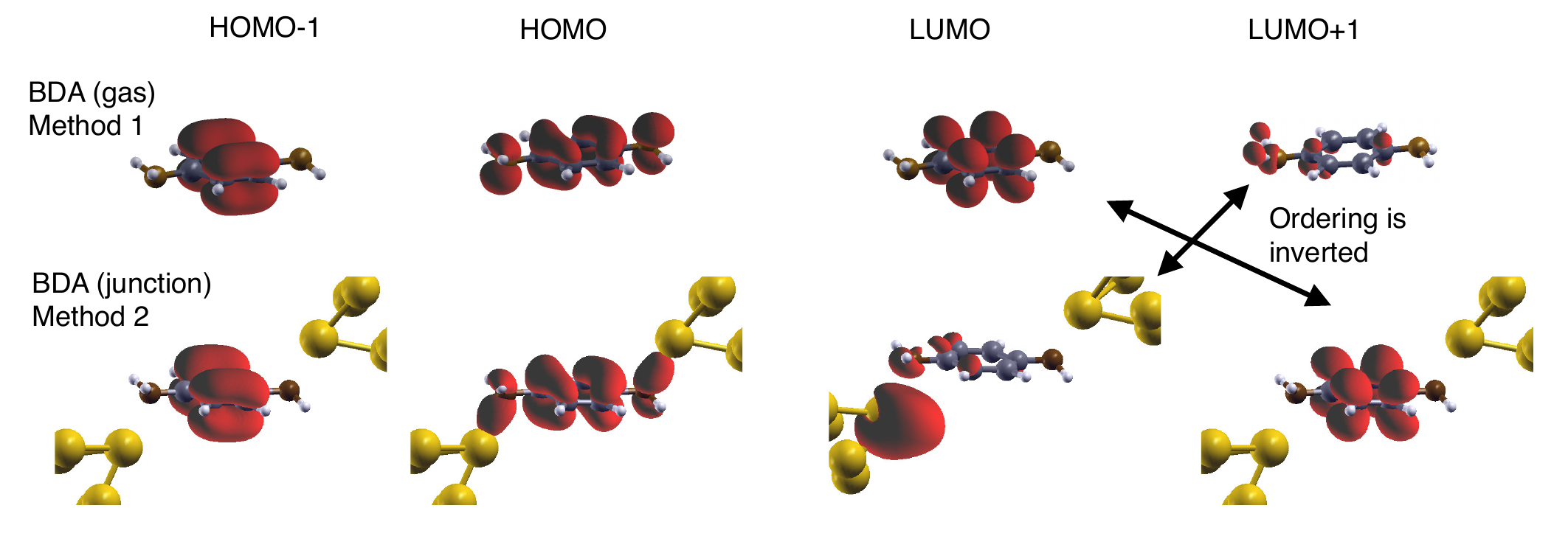}
\caption{Electronic density isosurfaces (red) of the HOMO-1, HOMO, LUMO and LUMO+1 molecular orbitals of BDA as obtained with the two traditional methods (see text).
The ordering of the LUMO and LUMO+1 is inverted in the two methods.
The localized MOs [HOMO-1 and LUMO (gas) or LUMO+1 (junction)] look remarkably similar for both methods.
In contrast, the HOMO and LUMO in the junction present a clear bonding with the leads and thus slightly differ from the corresponding MOs in gas-phase. 
Hydrogen, carbon, nitrogen and sulfur atoms are represented by white, grey, brown and green spheres, respectively.
}
\label{mos-bda}
\end{figure*}

\subsection{Computational details}

Our calculations are carried out within the DFT-Landauer framework.
The exchange-correlation energy is approximated using the PBE functional \cite{PBE}.
We use \textsc{ABINIT} \cite{abinit} for ground state calculations and \textsc{WanT} \cite{want,calzolari_want_2004} to construct Wannier functions and for conductance calculations.
All the results presented here are obtained by well converged calculations, using the same convergence parameters as in Ref.~\onlinecite{rangel_transport_2011}, which are consistent and in agreement with the literature.

\section{Results for benzene-diamine}

\subsection{BDA molecular orbitals}

In Fig.~\ref{mos-bda}, we show the molecular orbitals of BDA calculated with Methods 1 and 2. 
They are analogous to those found previously for instance using Method 1 \cite{ning_first-principles_2007}.
While the HOMO-1 molecular orbitals are very similar, the HOMO show non-negligible differences: the bonding character with the leads is more important when using Method 2, as indicated by the more pronounced lobes on the N atoms that point towards the gold adatoms.

We observe a close similarity between the LUMO from Method 1 and the LUMO+1 from Method 2, like if there were a change in the ordering of the corresponding eigenvalues between the two methods.
Notice that the energy difference between the LUMO and the LUMO+1 is $\sim$0.5 eV, so enough to exclude their degeneracy.
Vice versa, the LUMO+1 from Method 1 resembles the LUMO from Method 2 but there are some small differences: the bonding character with the leads is again more pronounced when using Method 2.
In fact, the corresponding density arises from a MLWF basis element which is localized on the gold-amino bond and not clearly identifiable as purely belonging to gold or to the molecule.
In this MO important differences are also found for the lobes on the benzene-ring: in Method 1, the lobes are mainly on the opposite C atoms along the molecule long axis; whereas, in Method 2, they are on the C atoms close to the Au adatom.

These differences will induce non-negligible differences in the PDOS analysis, as will see in the next section.

\subsection{PDOS and interpretation of the conductance}

\begin{figure}
\includegraphics[width=\columnwidth]{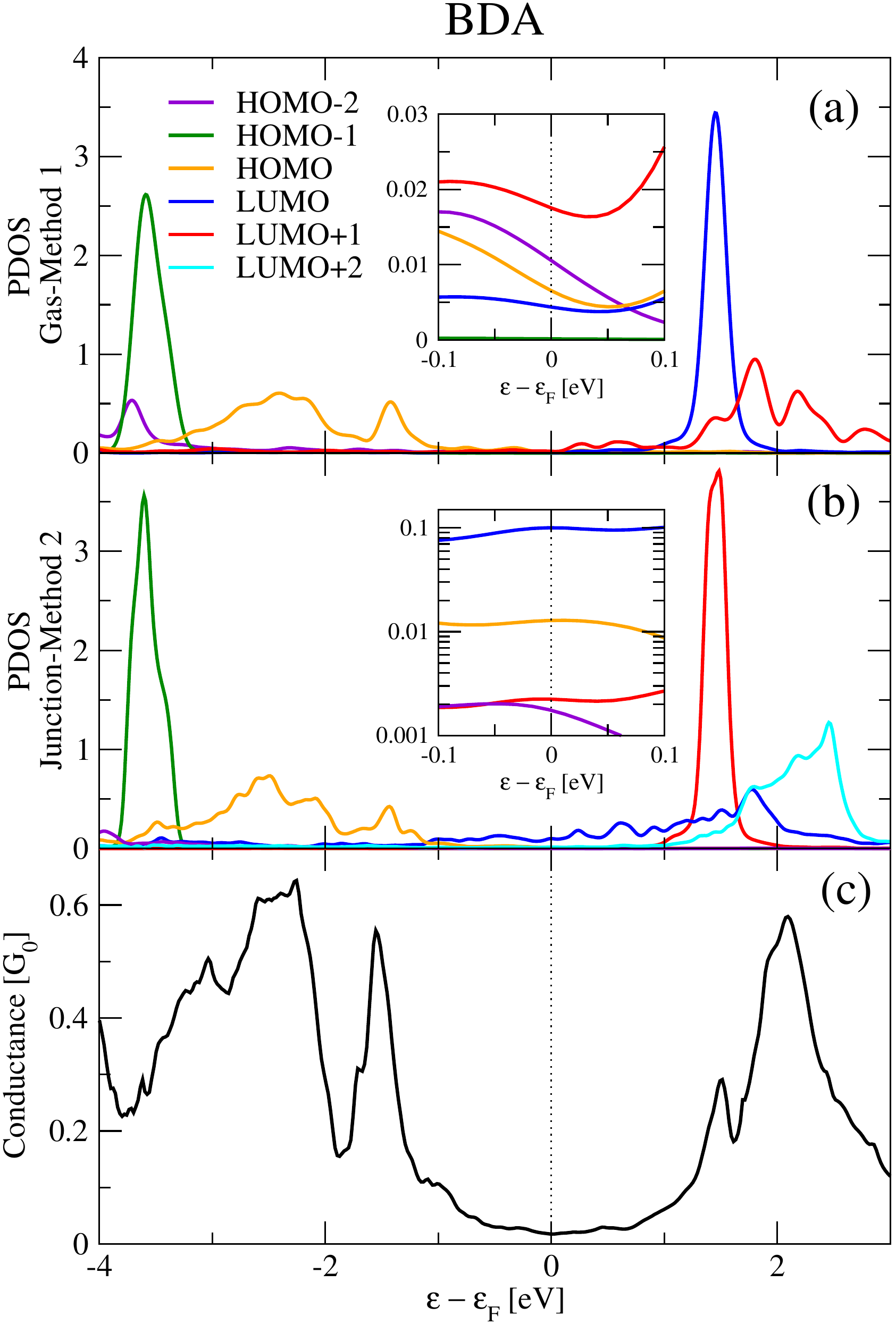}
\caption{Projected density of states (a,b) and conductance (c) of benzene-diamine (BDA).
The PDOS for the different molecular orbitals (from HOMO-2 to LUMO+1) have been obtained with (a) Method 1 and (b) Method 2 (see text).
The insets show a zoom on the PDOS zoom around the Fermi energy region.
Notice that, in the inset of (b), the PDOS is presented in logarithmic scale.}
\label{cond-pdos-bda}
\end{figure}

In Fig.~\ref{cond-pdos-bda}(c), we show the conductance of BDA calculated in the Landauer-DFT framework using the PBE approximation.
And, as it is usually done in literature for providing a physical interpretation of the conductance, we also present the PDOS as calculated using Methods 1 [Fig.~\ref{cond-pdos-bda}(a)] and 2 [Fig.~\ref{cond-pdos-bda}(b)].
The position and height of the main features are in very good agreement with previous work \cite{ning_first-principles_2007}.

The two PDOS look quite similar but with differences that can be associated to the already discussed discrepancies between MOs.
In particular, we observe the change in the ordering between the LUMO and the LUMO+1 from Method 1 to 2.
The PDOS onto non-hybridized MOs (HOMO-1 and LUMO/LUMO+1 in Method 1/2) look similar, whereas the PDOS onto the HOMO and LUMO+1/LUMO in Method 1/2 present differences, as expected from the MO plots.
Finally, the PDOS onto HOMO-2 seems to have more weight in Method 1 than in Method 2.

\begin{figure}
\includegraphics[width=\columnwidth]{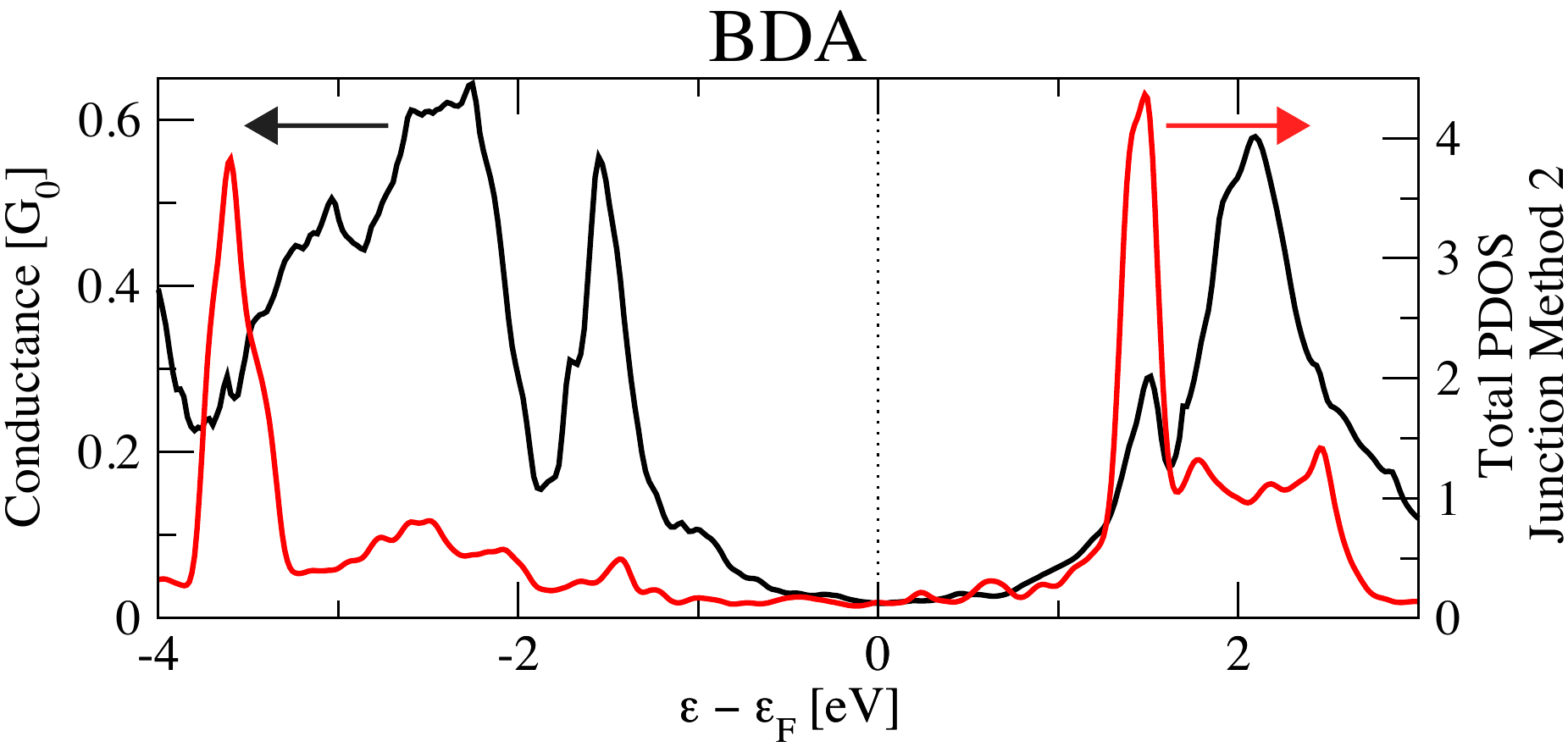}
\caption{Total molecular PDOS (red line) and conductance (black line) of BDA. The total molecular PDOS is the sum of the PDOS onto the MOs from HOMO-2 to LUMO+2 as obtained from Method 2. Note that the PDOS onto LUMO+2 is not shown in Fig.~\ref{cond-pdos-bda}. 
}
\label{cond-tpdos-bda}
\end{figure}

When interpreting the conductance profile, one can associate the small conductance peak at $\sim$1.5~eV with the intense LUMO and LUMO+1 PDOS peaks observed respectively in Methods 1 and 2.
The conductance structure arising at energies $>0$~eV with maximum at 2~eV could be correlated to the other unoccupied molecular orbital (LUMO+1 of BDA-gas \textit{alias} LUMO of BDA-junction), as well as the LUMO+2.
The peak in the conductance at $\sim$-1.5~eV could be related to the HOMO PDOS peak at $\sim$-1.4~eV, and so also the structure from -2 down to -3.8~eV.
The HOMO-1 and its PDOS peak at $\sim$-3.6~eV does not reflect in the conductance.
However, when performing a one-to-one comparison of the conductance with the total PDOS on the various MOs (Fig.~\ref{cond-tpdos-bda}), the relationship does not look that direct, even qualitatively.

We finally discuss the interpretation of the $0$-bias conductance.
Following one possible interpretation scheme very common in the literature, the zero-bias conductance appears on the tail of the conductance peak at $-1.5$~eV (HOMO), though the smallest peak at $+1.5$~eV (associated to the PDOS onto the LUMO/LUMO+1 from Method 1/2) is equally distant.
According to this interpretation, the zero-bias conductance is driven by the HOMO, though a contribution from the LUMO from Method 1 (\textit{alias} the LUMO+1 from Method 2) is expected.

These conclusions are contrasted by another approach which rather looks at the absolute values of the PDOS at the Fermi energy (see Fig.~\ref{cond-pdos-bda} insets showing zooms on the Fermi energy regions).
According to this scheme, the other unoccupied MO (the LUMO+1 from Method~1, \textit{alias} the LUMO from Method~2) drives the zero-bias conductance.
In fact, both methods agree on the fact that this MO (labeled differently) presents the largest PDOS value at the Fermi energy.
Nevertheless, its corresponding PDOS value at 0~eV disagrees by one order of magnitude:
from 0.1 in Method~2 to 0.02 in Method~1.
The next MO presenting an important PDOS value at the Fermi energy is the HOMO-2 from Method~1, with a value even not much smaller than the LUMO+1, implicating that the HOMO-2 has a certain weight on the zero-bias conductance.
However, this is the HOMO from Method~2 with a marked gap (from 0.1 to 0.01).
Both methods agree about the HOMO PDOS absolute value ($\sim$0.01) at $\epsilon=0$, probably by mere coincidence given the disagreements mentioned above.

Summarizing, when interpreting the BDA zero-bias conductance we are in front of 3 problems:
i) arbitrariness in the labeling of MOs (the LUMO in Method 1 becomes the LUMO+1 in Method 2, and vice versa);
ii) dependence on the method to identify the MOs;
iii) dependence on the interpreting approach.
Hence PDOS analyses of $\mathcal{G}(\epsilon=0)$ are affected by some ambiguity.

\begin{figure*}
\includegraphics[width=\textwidth]{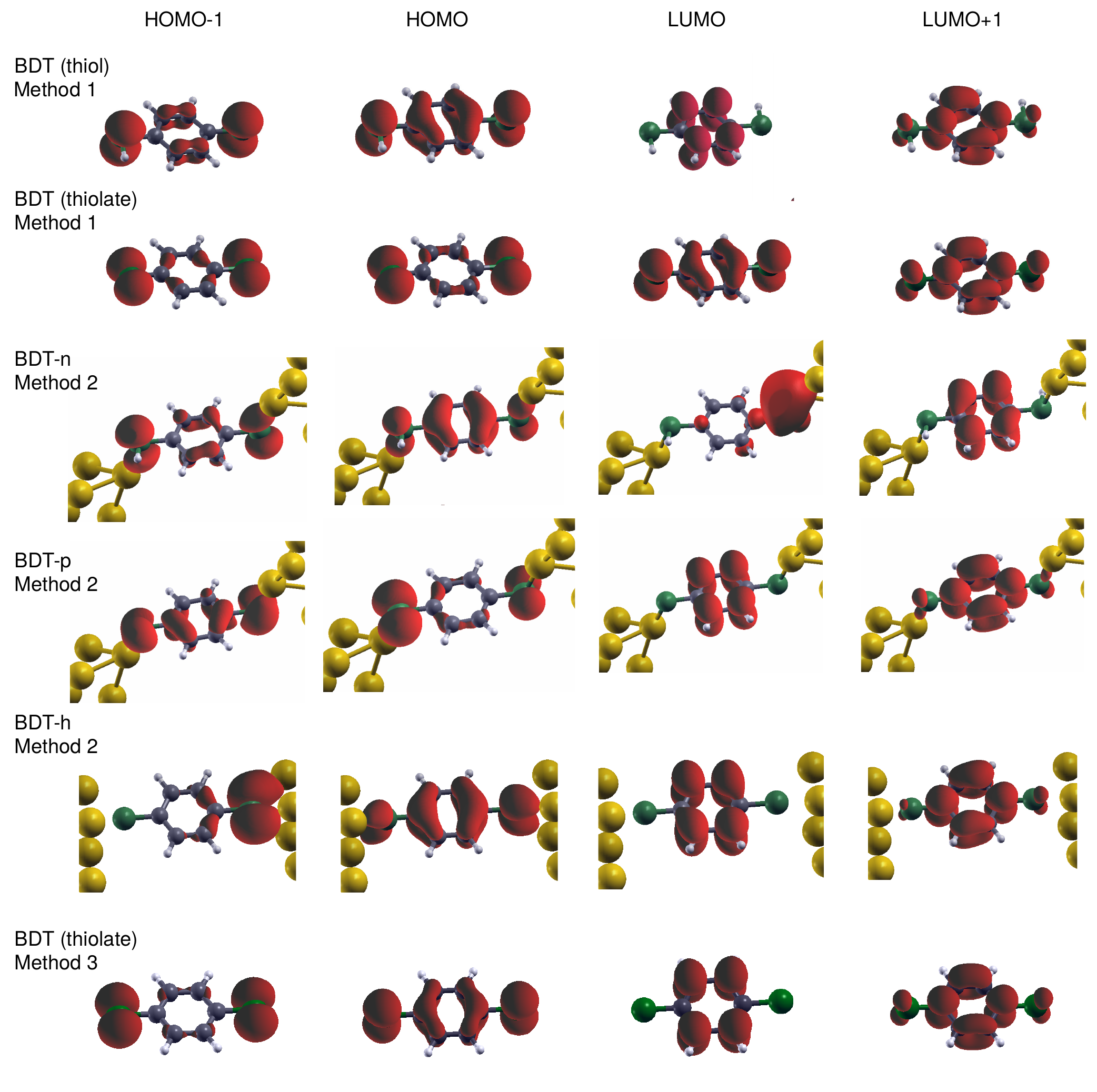}
\caption{Electronic density isosurfaces (red) of the HOMO-1, HOMO, LUMO and LUMO+1 molecular orbitals of BDT as obtained with the two traditional methods as well as with the new method (see text).
For Method 1, the dithiol and dithiolate molecules are considered. For Method 2, the different molecular junction geometries (BDT-n and BDT-p and BDT-h) are examined. For Method 3, a charge of $+0.5$~$e^-$ was added to the dithiolate molecule in order to account for the transfer of charge to the molecule from gold atoms in the BDT-h junction. The resulting orbitals are very similar to those obtained with Method 2 for BDT-h.
Hydrogen, carbon, nitrogen and sulfur atoms are represented by white, grey, brown and green spheres, respectively.
}
\label{mos-bdt}
\end{figure*}

\section{Results for benzene-dithiol}

\subsection{BDT molecular orbitals}

We now consider a more complex case: the benzene-dithiol~(BDT) gold junction.
Experimental and theoretical works concluded that the BDT-gold junction can be stable in several different atomic structures/geometries \cite{pontes_ab_2011,muller_effect_2006,french_structural_2013,french_large-scale_2012,souza_stretching_2014,nguyen_electric_2014}.
To account for different hybridizations and bonding motifs, three geometries are studied here:
the sulfur atom of the benzene-dithiol molecule can adsorb to an extra gold adatom without loosing the bound hydrogen atom (BDT-n);
the benzene-dithiol molecule can loose the hydrogen, thus becoming benzene-dithiolate, and bind its sulfur atom to an extra gold adatom in a pyramid structure (BDT-p);
or the benzene-dithiolate can bind to 3 equidistant gold atoms in the hollow structure (BDT-h).
These geometries are shown in Fig.~\ref{mos-bdt}.

In Fig.~\ref{mos-bdt}, we show the MOs of BDT calculated with different methods.
For Method~2, we show the molecular orbitals obtained for the 3 different junction geometries: BDT-n, BDT-p and BDT-h.
They are very similar to those obtained previously \cite{ThygesenJakobsen2005}, especially given the differences in the considered geometries.
In Ref.~\onlinecite{qian_calculating_2010}, an alternative set of MOs are shown for BDT-h, obtained within Method 2 by considering only the localized orbitals on the benzene molecule (excluding the S atoms).
For Method~1, we depict both the cases of benzene-dithiol and benzene-dithiolate.
The latter might better represent the BDT molecule in the BDT-p and BDT-h junctions where it looses a hydrogen atom before binding.
But this is not so straightforward: besides the effective chemical composition of the molecule in the junction, other chemical/physical effects, \textit{e.g.} saturation of bonds, transfer of charge, may be considered \cite{stadler_fermi_2006,stadler_fermi_2007,stadler_conformation_2010,kastlunger_charge_2013}.

We start by analyzing the MOs from Method 1.
The MOs for the dithiol and dithiolate molecules present a few similarities.
The LUMO+1 are similar in shape.
The HOMO of the dithiol molecule resembles to the LUMO of the dithiolate molecule, with an exchange of the ordering as already seen in BDA (see previous Section).
Nevertheless, other MOs strongly differ.
So, identification of MOs using Method 1 strongly depends on the molecule (dithiol vs. dithiolate).

Now we analyze the MOs obtained with Method~2.
We focus on BDT-p, the junction in which the interpretation of conductance using the PDOS is the most critical of all the cases considered here, as will be seen later.
The LUMO+1 from Method 2 looks very similar to the LUMO+1 from Method 1 for both the dithiol and dithiolate molecules, though with differences on the sulfur atom.
The LUMO from Method 2 corresponds to the LUMO from Method 1 for the dithiol molecule, but it has no correspondence to any MO from Method 1 for the dithiolate molecule.
On the other hand, the HOMO from Method 2 is similar to the HOMO from Method 1 for the dithiolate molecule, but 
it differs from all MOs from Method 1 for the dithiol molecule.
Finally, the MOs which look closer to the HOMO-1 from Method 2 are the HOMO from Method 1 for the dithiol molecule and the LUMO from Method 1 for the dithiolate molecule.

From the above discussion, it appears that no one-to-one correspondence can be established between the MOs obtained with the two methods nor between the MOs from Method 1 both for the dithiol and dithiolate isolated molecules.
The BDT-p MOs from Method~2 seem at half-way between the MOs from Method 1 for the dithiolate and dithiol molecules.

\subsection{PDOS and interpretation of the conductance}

\begin{figure}
\includegraphics[width=\columnwidth]{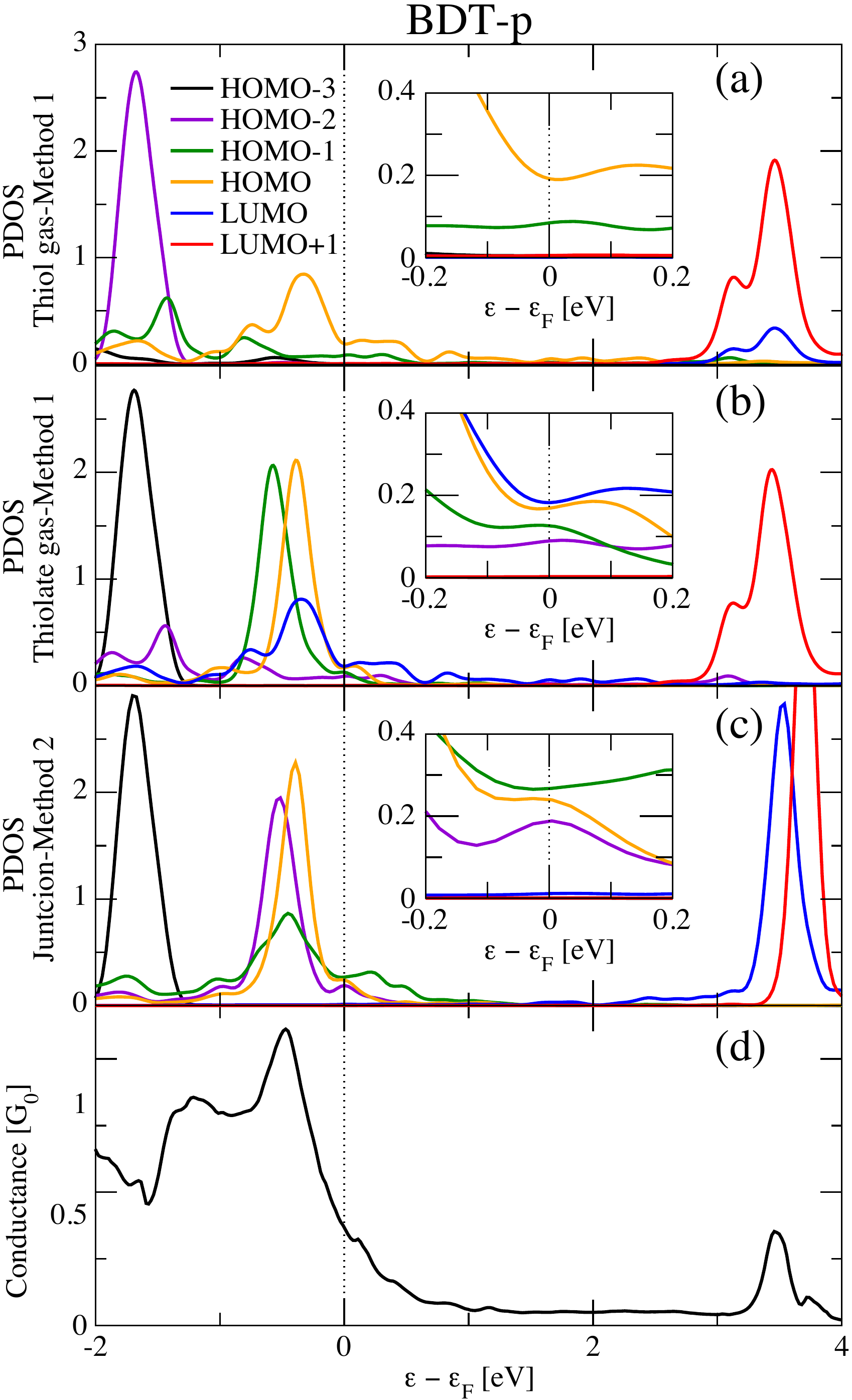}
\caption{Projected density of states (a,b,c) and conductance (d) of benzene-dithiol in the pyramid geometry (BDT-p).
The PDOS for the different molecular orbitals (from HOMO-3 to LUMO+1) have been obtained with (a) Method 1 based on the dithiol molecule, (b) Method 1 based on the dithiolate molecule, and (c) Method 2 (see text). 
The insets show a zoom on the PDOS zoom around the Fermi energy region.
}
\label{cond-pdos-bdt-p}
\end{figure}

We now move to the analysis of the most critical case (among the examples investigated here) regarding the interpretation of the conductance in terms of the PDOS: benzene-dithiol in the pyramid geometry (BDT-p).
In Fig.~\ref{cond-pdos-bdt-p} we present the Landauer-DFT conductance of BDT-p.
On top, we present also 3 different PDOS calculated following Method 1 (gas phase) and Method 2 (junction), for the former both the dithiol and dithiolate molecules are considered.

Without entering into all details, it is clear that the PDOS strongly depends on the method used to calculate it, reflecting previously seen differences in the MOs.
For instance, the zero-bias conductance seems dominated by the HOMO from Method~1 for the dithiol molecule, since the PDOS onto the HOMO is the closest to the Fermi level and it also provides the highest contribution at that level (see the inset), with minor contribution from the HOMO-1.
When using Method 1 for the dithiolate molecule, the zero-bias conductance seems equally driven by the HOMO and LUMO, with also some contribution from the HOMO-1 and the HOMO-2.
Finally, using Method 2, the HOMO-1, the HOMO and the HOMO-2 (in decreasing order) are the most important contributions at zero-bias .
Though some discrepancies can be ascribed to simple re-labeling of the same MO, one cannot pass over more important differences among the methods.

In conclusion, we could not find a rigorous definition of the MOs and associated PDOS for the BDT-p case when using the traditional methods.
As a consequence, the PDOS interpretation of the conductance does not rely on stable grounds.

\begin{figure}
\centerline{\includegraphics[width=\columnwidth]{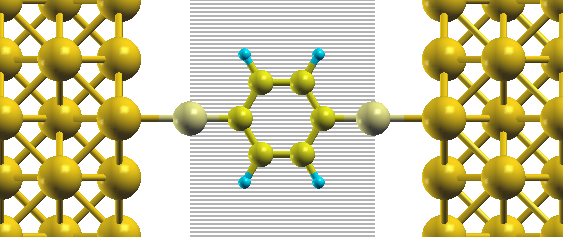}}
\caption{Scheme representing the integration volume (shadowed area passing through the two S atoms of the BDT-h junction) used for our Method~3.
}
\label{schema3}
\end{figure}

\section{New method for identifying molecular orbitals}

\subsection{Charged isolated molecules}

In order to reconcile the two main methods found in literature, that is, smooth their differences and solve the difficulties, we here propose a new approach which is based on an evolution of Method 1.
\begin{description}
\item[Method 3]
The $\phi^\mathrm{MO}_m(r)$ are chosen as the eigenfunctions of the Hamiltonian of the uncontacted, gas phase, isolated molecule, to which some charge is added accounting for metal-molecule charge-transfer. 
The same supercell is used as in the contacted molecule junction calculation, but removing the atoms of the leads. 
The added charge is calculated from a three steps procedure:
\vspace{-\topsep}
\begin{itemize}\itemsep2pt \parskip0pt \parsep0pt
\item[i)] the density $\rho(r)$ of the complete junction is computed;
\item[ii)] the density $\rho'(r)$ of the molecule is also calculated using the same geometry and simulation box as in the junction;
\item[iii)] the added charge is given by integrating $\rho(r)$$-$$\rho'(r)$ over the volume spanned by the molecule. 
For BDT-h, this volume is given by the region between two planes perpendicular to the S-S axis and passing through the two S atoms (see Fig.~\ref{schema3}).
\end{itemize}
\end{description}

The rationale behind our new method is to modify the electronic structure of the gas-phase isolated molecule with the purpose to account for the lead-molecule charge transfer.
Thus, the isolated molecule is placed into an environment closer to that one of the molecular-junction.
Previous studies~\cite{stadler_fermi_2006, stadler_conformation_2010} have already underlined the importance of the lead-molecule charge transfer and the significance of its role in transport properties of molecular junctions.
Here, it constitutes the basis for the construction of a new method of analysis.

\subsection{Application of the new method to BDT-h}

We apply our new method to the case of BDT-h (hollow geometry), which presents contradictory results using standard methods, as explained later.
According to our recipe, the extra charge to be added to BDT-thiolate to simulate the environment of the BDT-h junction was found to be $\sim$0.5~$e^-$.
However we observe that the modifications of the MOs are slightly affected by the precise value of the added charge, apart when the charge crosses integer values, $\rho$=0, 2, $\ldots$, of the electronic unit charge $e^-$, at the onset of the occupation of new levels.
The MOs found with this procedure are shown at the bottom of Fig.~\ref{mos-bdt}.
Remarkably, these MOs look now much more similar to the MOs found with Method 2 for BDT-h, as it can be clearly seen.
Furthermore, they present marked differences with the original Method 1 for dithiolate, and in some cases are even closer to Method 1 for the dithiol molecule.

\begin{figure}
\includegraphics[width=\columnwidth]{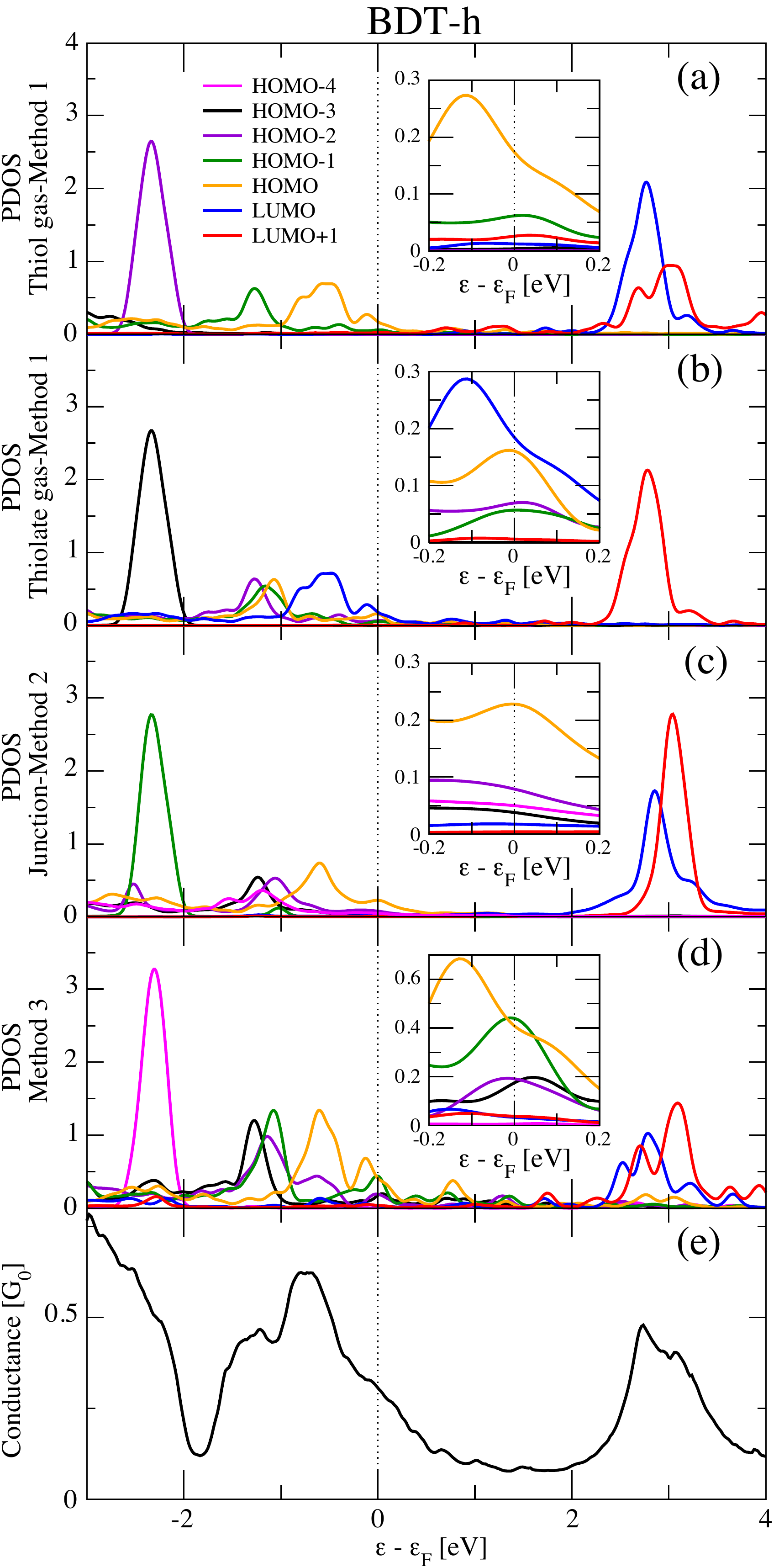}
\caption{Projected density of states (a,b,c,d) and conductance (e) of benzene-dithiol in the hollow geometry~(BDT-h).
The PDOS for the different molecular orbitals (from HOMO-3 to LUMO+1) have been obtained with (a) Method 1 based on the dithiol molecule, (b) Method 1 based on the dithiolate molecule, (c) Method 2, and (d) Method 3 (see text).
The insets show a zoom on the Fermi energy region.
}
\label{cond-pdos-bdt-h}
\end{figure}

Fig.~\ref{cond-pdos-bdt-h} shows the PDOS for the BDT-h junction calculated with the traditional methods (Method 1 for dithiol and dithiolate isolated molecules and Method 2 for a selected set of MLWFs for the BDT-h junction) and our new Method 3.
We focus on the PDOS around $\epsilon$=3~eV where traditional methods present the most important differences.
In that energy region, our new Method 3 provides an evident improvement.
Coming from a dithiolate isolated molecule, the PDOS from Method 3 is closer to the one from Method 1 for the dithiol molecule than for the dithiolate molecule, thus bridging the gap between the dithiol and dithiolate molecules.
Moreover, in this same energy region, when considering the relative height between the PDOS peaks of LUMO and LUMO+1, Method 3 evidently bridges the gap between Method 1 for the dithiol and dithiolate molecules and Method 2.
We can probably conclude the same also for the $\epsilon$=0~eV region, though restricting the discussion to the PDOS of the HOMO.
One can observe the evolution of the PDOS peak of the HOMO at the Fermi energy from Method 1 for the dithiol molecule, from Method 3 and from Method 2.
We can say that Method 3 is somehow successful in reconciling the traditional Methods 1 and 2.
However, we do not notice any other evident improvement.
We still find clear differences among the PDOS when zooming on the $\epsilon$=0~eV region (not shown).
The MO ordering problem continues to exist: the PDOS peak at $\sim$-2.5~eV from Method 3 is attributed to yet another MO, the HOMO-4.
The same ambiguous attribution remains for the PDOS of the intermediate HOMO orbitals.

We have tried Method 3 also on the more complex case of BDT-p.
The MOs from Method 3 (not shown) do not resemble to those from Method 2, and consequently we get no satisfactory results on the PDOS.
BDT-p continues to be an unsatisfactory case also for Method 3.
This is so probably because the metal-molecule charge transfer is not the only, or the main, parameter affecting the electronic structure of BDT-p, due to a may be higher metal-molecule coupling and hybridization.

In conclusion, Method 3 provides encouraging partial satisfactory results, in particular in reconciling the two traditional methods as in BDT-h. However, this is not general and not all problems are solved.
The metal-molecule charge transfer is not the only mechanism at play. 
One should probably take into account also the metal-molecule hybridization. This is not an easy task if the purpose is to keep the picture of an isolated molecule.

\section{Discussion: Further considerations on the PDOS analysis}

As discussed in the previous section, Method 3 aims at overcoming the drawbacks related to the identification of MOs using Method 1.
Instead, one could have explored the possibility to improve upon Method 2.
However, as we argue hereafter, this path appears to us less physically-grounded.
It actually opens even more fundamental questions on the implicit hypotheses at the basis of the interpretation of the conductance based on the PDOS, and raises further doubts on the validity of the whole procedure.

\subsection{Dependence of MOs and PDOS from the choice of Wannier Functions basis set}

At first sight, Method 2 (for which MOs originate from the junction) would seem more meaningful for studying the conductance.
However, it presents a severe drawback for which it seems very difficult to find a solution.
There is a certain arbitrariness in the criterion establishing the spatial limits of a molecule and thus the basis elements that will be considered as being ``localized on the molecule''.
For instance, there can be MLWFs localized on the molecule-lead bonds as we have pointed out for BDA.
It is then quite arbitrary to say whether they are localized on the molecule or on the leads.
This choice clearly affects the resulting submatrix, as well as the number and the shape of the MOs found after its diagonalization.

Intuitively, these basis elements should have an important effect on the junction conductance, so that it makes a lot of sense to keep them when generating the MOs. 
Coming back to the case of BDA, the most important PDOS at the Fermi energy was precisely the one associated to the MO presenting the major localization on the molecule-bond MLWF (i.e. the LUMO).
If we had discarded the latter from those ``localized on the molecule'', we would have excluded this important MO from the analysis of the zero-bias conductance.
It is actually reassuring that this MO also appeared when using Method 1, though labeled LUMO+1 due to the already discussed inverted ordering (see Fig.~\ref{mos-bda}) and it was also the most important PDOS at $\epsilon_F$.
But, at the same time, it shows that the exclusion of some MLWFs based on their localization may lead to very different interpretations starting from Method 1 or Method 2.

\begin{figure}
\includegraphics[width=\columnwidth]{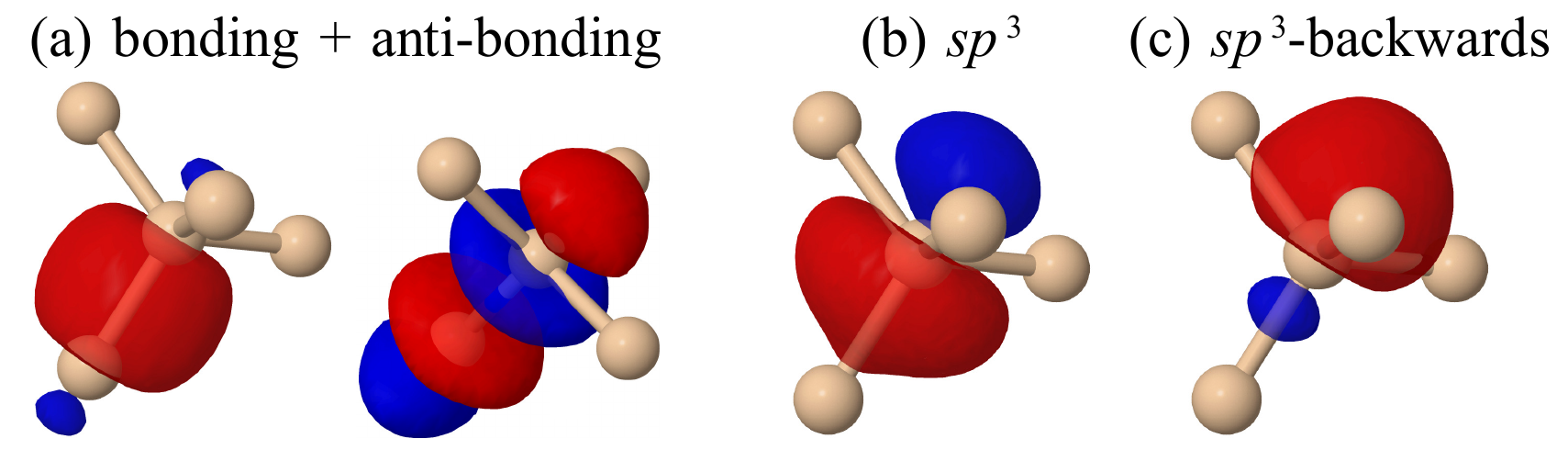}
\caption{
Using the Marzari-Vanderbilt algorithm, three different sets of Wannier functions (WFs) with comparable spread can be obtained for bulk silicon.
While a) bonding+antibonding and b) $sp^3$-forward are the most ``physical'' WFs, though not the most localized ones, c) $sp^3$-backwards are the maximally localized WFs.
}
\label{simlwf}
\end{figure}

A strategy to circumvent this drawback is to select a different set of Wannier functions (WFs), or any other localized basis set with elements presenting a well-defined localization (on the molecule or on the leads).
For instance, atom-centered basis sets would resolve this ambiguity, such as symmetry-adapted WFs \cite{sakuma_symmetry-adapted_2013}, WFs obtained from LCAO projections \cite{agapito_effective_2013}, or LCAO basis sets.
Furthermore, it is well-known that, in some cases, the Marzari-Vanderbilt\cite{marzari_maximally_1997} algorithm can lead to different sets of WFs.
For instance, silicon bulk presents at least 3 different sets of WFs with a similar degree of localization (as measured by the spread $S$).
When starting the Marzari-Vanderbilt algorithm from a random initial guess, there is a high probability to fall down into the global minimum ($S$=2.56~\AA$^2$) for which the lowest 8 MLWFs are of the $sp^3$-backward kind [Fig.~\ref{simlwf}(c)] which do not correspond to the real chemical orbitals.
It is obviously possible to obtain the 8 $sp^3$-forward WFs [Fig.~\ref{simlwf}(b)] which correspond to the physical chemical $sp^3$ orbitals, but at slightly higher local minimum ($S$=2.95~\AA$^2$).
Finally, the set of WFs with 4 bonding orbitals on one Si atom and 4 anti-bonding orbitals on the other atom [Fig.~\ref{simlwf}(a)] has a relatively large spread ($S$=5.09~\AA$^2$).
However, when performing the search of the MLWFs for the 4 valence states only, the minimum spread is obtained for a set containing the 4 bonding orbitals.

The previous discussion points to a possible ambiguity in Method 2 for identifying the MOs and hence in using the corresponding PDOS to interpret the conductance.
For a single junction, one may find several sets of WFs.
The one presenting the minimum spread (the most localized) does not necessarily correspond to the real physical situation, and this cannot be known \textit{a priori}.
The calculated conductance must and does not depend on the chosen basis set, provided the basis is complete and at convergence.
On the other hand, the submatrix of the junction Hamiltonian \textit{does} depend on the chosen basis set.
So do its eigenfunctions (which define the MOs) and the resultant PDOS.
Consequently, the physical interpretation of the conductance by the PDOS \textit{does} depend on the chosen WF or other basis set.
A basis-dependent interpretation method is questionable.

Starting from this point, we are led to ask even more fundamental questions:
Is the conductance really related to a MO, or a PDOS, or to some MOs and a total PDOS?
Before answering these questions, let us try to answer a question even further upstream.

\subsection{Is the conductance directly related to the full DOS?}

The conductance $\mathcal{G}(\epsilon)$ is certainly directly related to the electronic structure of the junction, \textit{i.e.} to both the electronic energies $\epsilon_n$ and wavefunctions $\phi_n(r)$ of the extended molecule.
Hence, there should also be a relationship to the total DOS, $\rho(\epsilon)$, though somehow indirect and not one-to-one. 
For instance, wherever $\rho(\epsilon)$=0 (there are no states available at that energy), the conductance $\mathcal{G}(\epsilon)$ must be also zero.
The reverse is not true: the conductance can be zero at energies where the total DOS is finite.
This can happen at energies associated to strongly localized wavefunctions, with zero spatial overlap among them, for example core states.
There can also be other factors beyond localization altering the direct relationship between $\mathcal{G}(\epsilon)$ and $\rho(\epsilon)$.
For instance, not all delocalized wavefunctions are good conducting channels \cite{lherbier_highly_2013}.
As a result, direct conclusions cannot be drawn from the inspection of the DOS only.

\subsection{Is the conductance related to some kind of PDOS?}

Whether the conductance is directly related to some kind of PDOS, be it onto a given MO or onto some MOs or even the total PDOS, is actually less obvious to answer than for the full DOS.
And so is the physical interpretation of the conductance based on such quantities.

Taking the example of BDT-h (Fig.~\ref{cond-pdos-bdt-h}), one can see that the conductance profile is qualitatively related to a total PDOS including the MOs which are close to the Fermi energy.
Nevertheless, it is not possible to observe a quantitative relationship between the conductance value and the total PDOS height.
This is more evident in the case of BDT-p (Fig.~\ref{cond-pdos-bdt-p}), one cannot explain why the conductance is larger at~$-1$~eV than at~$3.5$.
At below -1~eV, the agreement worsens even qualitatively.
In the case of BDA (see Fig.~\ref{cond-pdos-bda}), the relationship between the conductance and the total PDOS is even less evident.

This work has made it clear that the conductance analysis depends on a suitable choice of the MOs.
For this reason, the interpretation of the conductance in terms of the PDOS is quite questionable.
We should first provide an answer to the following fundamental questions:
\vspace{-\topsep}
\begin{itemize}\itemsep2pt \parskip0pt \parsep0pt
\item[i)] which set of MOs do physically represent the molecule in the junction?
\item[ii)] given the lead-molecule hybridization, are MOs obtained from an isolated molecule (i.e. from Methods 1 or 3) meaningful for analyzing a metal-molecule junction?
\item[iii)] are MOs obtained by diagonalizing a submatrix of the Hamiltonian (Method 2) physical, given the fact that they depend on the choice of basis set?
\end{itemize}

MOs identified as the eigenvectors of the gas phase, isolated Hamiltonian (Methods 1 and 3) have a physical meaning.
But, this is only true for the isolated molecule not necessarily for the junction.
For the latter, the eigenfunctions of the isolated molecule are nothing but yet another basis set (just like the atomic orbitals for a solid).
Furthermore, the actual choice of the molecule may not be unique (e.g., dithiol or dithiolate).
As for Method 2, an interpretation which depends on the chosen basis set (WFs, LCAO, Gaussians or Wavelets) cannot be considered physical.

We believe that a completely different direction should be taken in order to provide an answer to these questions.
What matters for a physical interpretation of the conductance is the full electronic structure of the extended molecule (containing also some layers of the leads).
Considering the extended molecule system needed to converge the conductance, which typically contains of the order of $10^2$ gold and $10^1$ molecule atoms, one can realize that the molecule does not even have such an important weight on the determination of the electronic structure of the junction.
Following these arguments, we can give the indication that a meaningful procedure to provide a physical interpretation of a junction conductance should rely on the wavefunctions and energies directly identified for the extended molecule electronic structure.
Thus, in order to provide a physical interpretation of the conductance, we believe that the local density-of-states (LDOS), a quantity independent from the basis set and directly built on the extended molecule wavefunctions and energies, is the most meaningful. 
Actually, we have already presented an application which uses the LDOS for the interpretation of the quantum transport conductance \cite{rangel_transport_2011}.

Regarding an interpretation of the molecular junction conductance rooted on the molecular PDOS, this work first tried to reconcile the two traditional methods (Methods 1 and 2) by introducing a new one (Method 3).
Some success was achieved in this direction, but we cannot consider the problem to be solved.
Further work is clearly needed.
However, our considerations led us to doubt that a fully satisfactory solution exists along this direction.

\section*{Conclusions}

Taking as examples two reference molecular junctions (benzene-diamine and benzene-dithiol between gold leads), we have investigated the interpretation of the conductance based on the projected density of states (PDOS) onto molecular orbitals.
The latter are usually identified following two procedures: diagonalization of Hamiltonian of the gas-phase isolated molecule (Method 1); 
and diagonalization of a submatrix of the junction Hamiltonian constructed by considering only basis elements localized on the molecule (Method 2).
We have shown that these two methods can lead to substantially different MOs and hence PDOS.
Furthermore, within Method 1, the PDOS depends on the isolated molecule chosen to represents the junction (\textit{e.g.} with or without dangling bonds); and,
within Method 2 the PDOS depends on the chosen basis set.
As a consequence, the analysis of the conductance based on the PDOS can lead to different, if not contrasting, conclusions.
This is particularly true for the analysis of the zero-bias conductance which can be found to be driven by, \textit{e.g.}, the LUMO in one method and the HOMO in another.
To go beyond these drawbacks, we proposed an alternative method (Method 3) as an improvement to Method 1.
This new method somehow reconciles Methods 1 and 2, but still presents problems which point to more fundamental questions.
An analysis of the conductance based on the PDOS seems not to rely on well established roots due to the arbitrariness in the identification of MOs.
Our proposal provided some indications toward possible solutions to the problem of interpreting the molecular junction conductance.

\section*{Acknowledgments}

We thank Pierre Darancet, Jeff Neaton and Xavier Blase for useful discussions.
GMR acknowledges the F.R.S.-FNRS for financial support.
Computational resources have been provided by the supercomputing facilities of the Universit\'e catholique de Louvain (CISM/UCL), by the Consortium des \'Equipements de Calcul Intensif en F\'ed\'eration Wallonie Bruxelles (C\'ECI), and by the French GENCI supercomputing center (Project i2012096-655).

\bibliography{qt}

\begin{thebibliography}{55}%
\makeatletter
\providecommand \@ifxundefined [1]{%
 \@ifx{#1\undefined}
}%
\providecommand \@ifnum [1]{%
 \ifnum #1\expandafter \@firstoftwo
 \else \expandafter \@secondoftwo
 \fi
}%
\providecommand \@ifx [1]{%
 \ifx #1\expandafter \@firstoftwo
 \else \expandafter \@secondoftwo
 \fi
}%
\providecommand \natexlab [1]{#1}%
\providecommand \enquote  [1]{``#1''}%
\providecommand \bibnamefont  [1]{#1}%
\providecommand \bibfnamefont [1]{#1}%
\providecommand \citenamefont [1]{#1}%
\providecommand \href@noop [0]{\@secondoftwo}%
\providecommand \href [0]{\begingroup \@sanitize@url \@href}%
\providecommand \@href[1]{\@@startlink{#1}\@@href}%
\providecommand \@@href[1]{\endgroup#1\@@endlink}%
\providecommand \@sanitize@url [0]{\catcode `\\12\catcode `\$12\catcode
  `\&12\catcode `\#12\catcode `\^12\catcode `\_12\catcode `\%12\relax}%
\providecommand \@@startlink[1]{}%
\providecommand \@@endlink[0]{}%
\providecommand \url  [0]{\begingroup\@sanitize@url \@url }%
\providecommand \@url [1]{\endgroup\@href {#1}{\urlprefix }}%
\providecommand \urlprefix  [0]{URL }%
\providecommand \Eprint [0]{\href }%
\providecommand \doibase [0]{http://dx.doi.org/}%
\providecommand \selectlanguage [0]{\@gobble}%
\providecommand \bibinfo  [0]{\@secondoftwo}%
\providecommand \bibfield  [0]{\@secondoftwo}%
\providecommand \translation [1]{[#1]}%
\providecommand \BibitemOpen [0]{}%
\providecommand \bibitemStop [0]{}%
\providecommand \bibitemNoStop [0]{.\EOS\space}%
\providecommand \EOS [0]{\spacefactor3000\relax}%
\providecommand \BibitemShut  [1]{\csname bibitem#1\endcsname}%
\let\auto@bib@innerbib\@empty
\bibitem [{\citenamefont {Reed}\ \emph {et~al.}(1997)\citenamefont {Reed},
  \citenamefont {Zhou}, \citenamefont {Muller}, \citenamefont {Burgin},\ and\
  \citenamefont {Tour}}]{reed_conductance_1997}%
  \BibitemOpen
  \bibfield  {author} {\bibinfo {author} {\bibfnamefont {M.~A.}\ \bibnamefont
  {Reed}}, \bibinfo {author} {\bibfnamefont {C.}~\bibnamefont {Zhou}}, \bibinfo
  {author} {\bibfnamefont {C.~J.}\ \bibnamefont {Muller}}, \bibinfo {author}
  {\bibfnamefont {T.~P.}\ \bibnamefont {Burgin}}, \ and\ \bibinfo {author}
  {\bibfnamefont {J.~M.}\ \bibnamefont {Tour}},\ }\href@noop {} {\bibfield
  {journal} {\bibinfo  {journal} {Science}\ }\textbf {\bibinfo {volume}
  {278}},\ \bibinfo {pages} {252} (\bibinfo {year} {1997})}\BibitemShut
  {NoStop}%
\bibitem [{\citenamefont {Xu}\ and\ \citenamefont
  {Tao}(2003)}]{xu_measurement_2003}%
  \BibitemOpen
  \bibfield  {author} {\bibinfo {author} {\bibfnamefont {B.}~\bibnamefont
  {Xu}}\ and\ \bibinfo {author} {\bibfnamefont {N.~J.}\ \bibnamefont {Tao}},\
  }\href@noop {} {\bibfield  {journal} {\bibinfo  {journal} {Science}\ }\textbf
  {\bibinfo {volume} {301}},\ \bibinfo {pages} {1221} (\bibinfo {year}
  {2003})}\BibitemShut {NoStop}%
\bibitem [{\citenamefont {Park}\ \emph {et~al.}(2000)\citenamefont {Park},
  \citenamefont {Park}, \citenamefont {Lim}, \citenamefont {Anderson},
  \citenamefont {Alivisatos},\ and\ \citenamefont
  {McEuen}}]{park_nanomechanical_2000}%
  \BibitemOpen
  \bibfield  {author} {\bibinfo {author} {\bibfnamefont {H.}~\bibnamefont
  {Park}}, \bibinfo {author} {\bibfnamefont {J.}~\bibnamefont {Park}}, \bibinfo
  {author} {\bibfnamefont {A.~K.~L.}\ \bibnamefont {Lim}}, \bibinfo {author}
  {\bibfnamefont {E.~H.}\ \bibnamefont {Anderson}}, \bibinfo {author}
  {\bibfnamefont {A.~P.}\ \bibnamefont {Alivisatos}}, \ and\ \bibinfo {author}
  {\bibfnamefont {P.~L.}\ \bibnamefont {McEuen}},\ }\href@noop {} {\bibfield
  {journal} {\bibinfo  {journal} {Nature}\ }\textbf {\bibinfo {volume} {407}},\
  \bibinfo {pages} {57} (\bibinfo {year} {2000})}\BibitemShut {NoStop}%
\bibitem [{\citenamefont {Venkataraman}\ \emph {et~al.}(2006)\citenamefont
  {Venkataraman}, \citenamefont {Klare}, \citenamefont {Tam}, \citenamefont
  {Nucko~lls}, \citenamefont {Hybertsen},\ and\ \citenamefont
  {Steigerwald}}]{venkataraman_single-molecule_2006}%
  \BibitemOpen
  \bibfield  {author} {\bibinfo {author} {\bibfnamefont {L.}~\bibnamefont
  {Venkataraman}}, \bibinfo {author} {\bibfnamefont {J.~E.}\ \bibnamefont
  {Klare}}, \bibinfo {author} {\bibfnamefont {I.~W.}\ \bibnamefont {Tam}},
  \bibinfo {author} {\bibfnamefont {C.}~\bibnamefont {Nucko~lls}}, \bibinfo
  {author} {\bibfnamefont {M.~S.}\ \bibnamefont {Hybertsen}}, \ and\ \bibinfo
  {author} {\bibfnamefont {M.~L.}\ \bibnamefont {Steigerwald}},\ }\href@noop {}
  {\bibfield  {journal} {\bibinfo  {journal} {Nano Lett.}\ }\textbf {\bibinfo
  {volume} {6}},\ \bibinfo {pages} {458} (\bibinfo {year} {2006})}\BibitemShut
  {NoStop}%
\bibitem [{\citenamefont {Datta}(1995)}]{Datta}%
  \BibitemOpen
  \bibfield  {author} {\bibinfo {author} {\bibfnamefont {S.}~\bibnamefont
  {Datta}},\ }\href@noop {} {\emph {\bibinfo {title} {Electronic transport in
  mesoscopic systems}}}\ (\bibinfo  {publisher} {Cambridge university press},\
  \bibinfo {address} {Cambridge},\ \bibinfo {year} {1995})\BibitemShut
  {NoStop}%
\bibitem [{\citenamefont {{Di Ventra}}(2008)}]{DiVentra}%
  \BibitemOpen
  \bibfield  {author} {\bibinfo {author} {\bibfnamefont {M.}~\bibnamefont {{Di
  Ventra}}},\ }\href@noop {} {\emph {\bibinfo {title} {Electrical transport in
  nanoscale systems}}}\ (\bibinfo  {publisher} {Cambridge university press},\
  \bibinfo {address} {Cambridge},\ \bibinfo {year} {2008})\BibitemShut
  {NoStop}%
\bibitem [{\citenamefont {Kubo}(1958)}]{Kubo}%
  \BibitemOpen
  \bibfield  {author} {\bibinfo {author} {\bibfnamefont {R.}~\bibnamefont
  {Kubo}},\ }\href@noop {} {\bibfield  {journal} {\bibinfo  {journal} {J. Phys.
  Soc. Japan}\ }\textbf {\bibinfo {volume} {12}},\ \bibinfo {pages} {570}
  (\bibinfo {year} {1958})}\BibitemShut {NoStop}%
\bibitem [{\citenamefont {Greenwood}(1958)}]{Greenwood}%
  \BibitemOpen
  \bibfield  {author} {\bibinfo {author} {\bibfnamefont {D.~A.}\ \bibnamefont
  {Greenwood}},\ }\href@noop {} {\bibfield  {journal} {\bibinfo  {journal}
  {Proc. Phys. Soc.}\ }\textbf {\bibinfo {volume} {71}},\ \bibinfo {pages}
  {585} (\bibinfo {year} {1958})}\BibitemShut {NoStop}%
\bibitem [{\citenamefont {Landauer}(1957)}]{Landauer}%
  \BibitemOpen
  \bibfield  {author} {\bibinfo {author} {\bibfnamefont {R.}~\bibnamefont
  {Landauer}},\ }\href@noop {} {\bibfield  {journal} {\bibinfo  {journal} {IBM
  Journal of Research and Development}\ }\textbf {\bibinfo {volume} {1}},\
  \bibinfo {pages} {223} (\bibinfo {year} {1957})}\BibitemShut {NoStop}%
\bibitem [{\citenamefont {Keldysh}(1965)}]{Keldysh}%
  \BibitemOpen
  \bibfield  {author} {\bibinfo {author} {\bibfnamefont {L.~V.}\ \bibnamefont
  {Keldysh}},\ }\href@noop {} {\bibfield  {journal} {\bibinfo  {journal} {Sov.
  Phys. JETP}\ }\textbf {\bibinfo {volume} {20}},\ \bibinfo {pages} {1018}
  (\bibinfo {year} {1965})}\BibitemShut {NoStop}%
\bibitem [{\citenamefont {Danielewicz}(1984)}]{Danielewicz}%
  \BibitemOpen
  \bibfield  {author} {\bibinfo {author} {\bibfnamefont {P.}~\bibnamefont
  {Danielewicz}},\ }\href@noop {} {\bibfield  {journal} {\bibinfo  {journal}
  {Ann. Phys.}\ }\textbf {\bibinfo {volume} {152}},\ \bibinfo {pages} {239}
  (\bibinfo {year} {1984})}\BibitemShut {NoStop}%
\bibitem [{\citenamefont {Rammer}\ and\ \citenamefont
  {Smith}(1986)}]{RammerSmith}%
  \BibitemOpen
  \bibfield  {author} {\bibinfo {author} {\bibfnamefont {J.}~\bibnamefont
  {Rammer}}\ and\ \bibinfo {author} {\bibfnamefont {H.}~\bibnamefont {Smith}},\
  }\href@noop {} {\bibfield  {journal} {\bibinfo  {journal} {Rev. Mod. Phys}\
  }\textbf {\bibinfo {volume} {58}},\ \bibinfo {pages} {323} (\bibinfo {year}
  {1986})}\BibitemShut {NoStop}%
\bibitem [{\citenamefont {Thygesen}\ and\ \citenamefont
  {Jakobsen}(2005)}]{ThygesenJakobsen2005}%
  \BibitemOpen
  \bibfield  {author} {\bibinfo {author} {\bibfnamefont {K.~S.}\ \bibnamefont
  {Thygesen}}\ and\ \bibinfo {author} {\bibfnamefont {K.~W.}\ \bibnamefont
  {Jakobsen}},\ }\href@noop {} {\bibfield  {journal} {\bibinfo  {journal}
  {Chem. Phys.}\ }\textbf {\bibinfo {volume} {319}},\ \bibinfo {pages} {111}
  (\bibinfo {year} {2005})}\BibitemShut {NoStop}%
\bibitem [{\citenamefont {Nitzan}\ and\ \citenamefont
  {Ratner}(2013)}]{NitzanRatner}%
  \BibitemOpen
  \bibfield  {author} {\bibinfo {author} {\bibfnamefont {A.}~\bibnamefont
  {Nitzan}}\ and\ \bibinfo {author} {\bibfnamefont {M.~A.}\ \bibnamefont
  {Ratner}},\ }\href@noop {} {\bibfield  {journal} {\bibinfo  {journal}
  {Science}\ }\textbf {\bibinfo {volume} {300}},\ \bibinfo {pages} {1384}
  (\bibinfo {year} {2013})}\BibitemShut {NoStop}%
\bibitem [{\citenamefont {Toher}\ and\ \citenamefont
  {Sanvito}(2008)}]{toher_effects_2008}%
  \BibitemOpen
  \bibfield  {author} {\bibinfo {author} {\bibfnamefont {C.}~\bibnamefont
  {Toher}}\ and\ \bibinfo {author} {\bibfnamefont {S.}~\bibnamefont
  {Sanvito}},\ }\href@noop {} {\bibfield  {journal} {\bibinfo  {journal} {Phys.
  Rev. B}\ }\textbf {\bibinfo {volume} {77}},\ \bibinfo {pages} {155402}
  (\bibinfo {year} {2008})}\BibitemShut {NoStop}%
\bibitem [{\citenamefont {Pontes}\ \emph {et~al.}(2011)\citenamefont {Pontes},
  \citenamefont {Rocha}, \citenamefont {Sanvito}, \citenamefont {Fazzio},\ and\
  \citenamefont {da~Silva}}]{pontes_ab_2011}%
  \BibitemOpen
  \bibfield  {author} {\bibinfo {author} {\bibfnamefont {R.~B.}\ \bibnamefont
  {Pontes}}, \bibinfo {author} {\bibfnamefont {A.~R.}\ \bibnamefont {Rocha}},
  \bibinfo {author} {\bibfnamefont {S.}~\bibnamefont {Sanvito}}, \bibinfo
  {author} {\bibfnamefont {A.}~\bibnamefont {Fazzio}}, \ and\ \bibinfo {author}
  {\bibfnamefont {A.~J.~R.}\ \bibnamefont {da~Silva}},\ }\href@noop {}
  {\bibfield  {journal} {\bibinfo  {journal} {ACS Nano}\ }\textbf {\bibinfo
  {volume} {5}},\ \bibinfo {pages} {795} (\bibinfo {year} {2011})}\BibitemShut
  {NoStop}%
\bibitem [{\citenamefont {Ferretti}\ \emph {et~al.}(2012)\citenamefont
  {Ferretti}, \citenamefont {Mallia}, \citenamefont {Martin-Samos},
  \citenamefont {Bussi}, \citenamefont {Ruini}, \citenamefont {Montanari},\
  and\ \citenamefont {Harrison}}]{ferretti_ab_2012}%
  \BibitemOpen
  \bibfield  {author} {\bibinfo {author} {\bibfnamefont {A.}~\bibnamefont
  {Ferretti}}, \bibinfo {author} {\bibfnamefont {G.}~\bibnamefont {Mallia}},
  \bibinfo {author} {\bibfnamefont {L.}~\bibnamefont {Martin-Samos}}, \bibinfo
  {author} {\bibfnamefont {G.}~\bibnamefont {Bussi}}, \bibinfo {author}
  {\bibfnamefont {A.}~\bibnamefont {Ruini}}, \bibinfo {author} {\bibfnamefont
  {B.}~\bibnamefont {Montanari}}, \ and\ \bibinfo {author} {\bibfnamefont
  {N.~M.}\ \bibnamefont {Harrison}},\ }\href@noop {} {\bibfield  {journal}
  {\bibinfo  {journal} {Phys. Rev. B}\ }\textbf {\bibinfo {volume} {85}},\
  \bibinfo {pages} {235105} (\bibinfo {year} {2012})}\BibitemShut {NoStop}%
\bibitem [{\citenamefont {Quek}\ \emph {et~al.}(2007)\citenamefont {Quek},
  \citenamefont {Venkataraman}, \citenamefont {Choi}, \citenamefont {Louie},
  \citenamefont {Hybertsen},\ and\ \citenamefont
  {Neaton}}]{quek_aminegold_2007}%
  \BibitemOpen
  \bibfield  {author} {\bibinfo {author} {\bibfnamefont {S.~Y.}\ \bibnamefont
  {Quek}}, \bibinfo {author} {\bibfnamefont {L.}~\bibnamefont {Venkataraman}},
  \bibinfo {author} {\bibfnamefont {H.~J.~n.}\ \bibnamefont {Choi}}, \bibinfo
  {author} {\bibfnamefont {S.~G.}\ \bibnamefont {Louie}}, \bibinfo {author}
  {\bibfnamefont {M.~S.}\ \bibnamefont {Hybertsen}}, \ and\ \bibinfo {author}
  {\bibfnamefont {J.~B.}\ \bibnamefont {Neaton}},\ }\href@noop {} {\bibfield
  {journal} {\bibinfo  {journal} {Nano Lett.}\ }\textbf {\bibinfo {volume}
  {7}},\ \bibinfo {pages} {3477} (\bibinfo {year} {2007})}\BibitemShut
  {NoStop}%
\bibitem [{\citenamefont {Darancet}\ \emph {et~al.}(2012)\citenamefont
  {Darancet}, \citenamefont {Widawsky}, \citenamefont {Choi}, \citenamefont
  {Venkataraman},\ and\ \citenamefont {Neaton}}]{darancet_quantitative_2012}%
  \BibitemOpen
  \bibfield  {author} {\bibinfo {author} {\bibfnamefont {P.}~\bibnamefont
  {Darancet}}, \bibinfo {author} {\bibfnamefont {J.~R.}\ \bibnamefont
  {Widawsky}}, \bibinfo {author} {\bibfnamefont {H.~J.}\ \bibnamefont {Choi}},
  \bibinfo {author} {\bibfnamefont {L.}~\bibnamefont {Venkataraman}}, \ and\
  \bibinfo {author} {\bibfnamefont {J.~B.}\ \bibnamefont {Neaton}},\
  }\href@noop {} {\bibfield  {journal} {\bibinfo  {journal} {Nano Lett.}\
  }\textbf {\bibinfo {volume} {12}},\ \bibinfo {pages} {6250} (\bibinfo {year}
  {2012})}\BibitemShut {NoStop}%
\bibitem [{\citenamefont {Dell'Angela}\ \emph {et~al.}(2010)\citenamefont
  {Dell'Angela}, \citenamefont {Kladnik}, \citenamefont {Cossaro},
  \citenamefont {Verdini}, \citenamefont {Kamenetska}, \citenamefont {Tamblyn},
  \citenamefont {Quek}, \citenamefont {Neaton}, \citenamefont {Cvetko},
  \citenamefont {Morgante},\ and\ \citenamefont
  {Venkataraman}}]{dellangela_relating_2010}%
  \BibitemOpen
  \bibfield  {author} {\bibinfo {author} {\bibfnamefont {M.}~\bibnamefont
  {Dell'Angela}}, \bibinfo {author} {\bibfnamefont {G.}~\bibnamefont
  {Kladnik}}, \bibinfo {author} {\bibfnamefont {A.}~\bibnamefont {Cossaro}},
  \bibinfo {author} {\bibfnamefont {A.}~\bibnamefont {Verdini}}, \bibinfo
  {author} {\bibfnamefont {M.}~\bibnamefont {Kamenetska}}, \bibinfo {author}
  {\bibfnamefont {I.}~\bibnamefont {Tamblyn}}, \bibinfo {author} {\bibfnamefont
  {S.~Y.}\ \bibnamefont {Quek}}, \bibinfo {author} {\bibfnamefont {J.~B.}\
  \bibnamefont {Neaton}}, \bibinfo {author} {\bibfnamefont {D.}~\bibnamefont
  {Cvetko}}, \bibinfo {author} {\bibfnamefont {A.}~\bibnamefont {Morgante}}, \
  and\ \bibinfo {author} {\bibfnamefont {L.}~\bibnamefont {Venkataraman}},\
  }\href@noop {} {\bibfield  {journal} {\bibinfo  {journal} {Nano Lett.}\
  }\textbf {\bibinfo {volume} {10}},\ \bibinfo {pages} {2470} (\bibinfo {year}
  {2010})}\BibitemShut {NoStop}%
\bibitem [{\citenamefont {Cehovin}\ \emph {et~al.}(2008)\citenamefont
  {Cehovin}, \citenamefont {Mera}, \citenamefont {Jensen}, \citenamefont
  {Stokbro},\ and\ \citenamefont {Pedersen}}]{cehovin_role_2008}%
  \BibitemOpen
  \bibfield  {author} {\bibinfo {author} {\bibfnamefont {A.}~\bibnamefont
  {Cehovin}}, \bibinfo {author} {\bibfnamefont {H.}~\bibnamefont {Mera}},
  \bibinfo {author} {\bibfnamefont {J.~H.}\ \bibnamefont {Jensen}}, \bibinfo
  {author} {\bibfnamefont {K.}~\bibnamefont {Stokbro}}, \ and\ \bibinfo
  {author} {\bibfnamefont {T.~.~B.}\ \bibnamefont {Pedersen}},\ }\href@noop {}
  {\bibfield  {journal} {\bibinfo  {journal} {Phys. Rev. B}\ }\textbf {\bibinfo
  {volume} {77}},\ \bibinfo {pages} {195432} (\bibinfo {year}
  {2008})}\BibitemShut {NoStop}%
\bibitem [{\citenamefont {Strange}\ \emph {et~al.}(2011)\citenamefont
  {Strange}, \citenamefont {Rostgaard}, \citenamefont {Hakkinen},\ and\
  \citenamefont {Thygesen}}]{strange}%
  \BibitemOpen
  \bibfield  {author} {\bibinfo {author} {\bibfnamefont {M.}~\bibnamefont
  {Strange}}, \bibinfo {author} {\bibfnamefont {C.}~\bibnamefont {Rostgaard}},
  \bibinfo {author} {\bibfnamefont {H.}~\bibnamefont {Hakkinen}}, \ and\
  \bibinfo {author} {\bibfnamefont {K.~S.}\ \bibnamefont {Thygesen}},\
  }\href@noop {} {\bibfield  {journal} {\bibinfo  {journal} {Phys. Rev. B}\ ,\
  \bibinfo {pages} {115108}} (\bibinfo {year} {2011})}\BibitemShut {NoStop}%
\bibitem [{\citenamefont {Rangel}\ \emph {et~al.}(2011)\citenamefont {Rangel},
  \citenamefont {Ferretti}, \citenamefont {Trevisanutto}, \citenamefont
  {Olevano},\ and\ \citenamefont {Rignanese}}]{rangel_transport_2011}%
  \BibitemOpen
  \bibfield  {author} {\bibinfo {author} {\bibfnamefont {T.}~\bibnamefont
  {Rangel}}, \bibinfo {author} {\bibfnamefont {A.}~\bibnamefont {Ferretti}},
  \bibinfo {author} {\bibfnamefont {P.~E.}\ \bibnamefont {Trevisanutto}},
  \bibinfo {author} {\bibfnamefont {V.}~\bibnamefont {Olevano}}, \ and\
  \bibinfo {author} {\bibfnamefont {G.-M.}\ \bibnamefont {Rignanese}},\
  }\href@noop {} {\bibfield  {journal} {\bibinfo  {journal} {Phys. Rev. B}\
  }\textbf {\bibinfo {volume} {84}},\ \bibinfo {pages} {045426} (\bibinfo
  {year} {2011})}\BibitemShut {NoStop}%
\bibitem [{\citenamefont {Stadler}\ and\ \citenamefont
  {Jacobsen}(2006)}]{stadler_fermi_2006}%
  \BibitemOpen
  \bibfield  {author} {\bibinfo {author} {\bibfnamefont {R.}~\bibnamefont
  {Stadler}}\ and\ \bibinfo {author} {\bibfnamefont {K.~W.}\ \bibnamefont
  {Jacobsen}},\ }\href@noop {} {\bibfield  {journal} {\bibinfo  {journal}
  {Phys. Rev. B}\ }\textbf {\bibinfo {volume} {74}},\ \bibinfo {pages} {161405}
  (\bibinfo {year} {2006})}\BibitemShut {NoStop}%
\bibitem [{\citenamefont {Stadler}(2007)}]{stadler_fermi_2007}%
  \BibitemOpen
  \bibfield  {author} {\bibinfo {author} {\bibfnamefont {R.}~\bibnamefont
  {Stadler}},\ }\href@noop {} {\bibfield  {journal} {\bibinfo  {journal} {J.
  Phys Conf. Ser.}\ }\textbf {\bibinfo {volume} {61}},\ \bibinfo {pages} {1097}
  (\bibinfo {year} {2007})}\BibitemShut {NoStop}%
\bibitem [{\citenamefont {Stadler}(2010)}]{stadler_conformation_2010}%
  \BibitemOpen
  \bibfield  {author} {\bibinfo {author} {\bibfnamefont {R.}~\bibnamefont
  {Stadler}},\ }\href@noop {} {\bibfield  {journal} {\bibinfo  {journal} {Phys.
  Rev. B}\ }\textbf {\bibinfo {volume} {81}},\ \bibinfo {pages} {165429}
  (\bibinfo {year} {2010})}\BibitemShut {NoStop}%
\bibitem [{\citenamefont {Kastlunger}\ and\ \citenamefont
  {Stadler}(2013)}]{kastlunger_charge_2013}%
  \BibitemOpen
  \bibfield  {author} {\bibinfo {author} {\bibfnamefont {G.}~\bibnamefont
  {Kastlunger}}\ and\ \bibinfo {author} {\bibfnamefont {R.}~\bibnamefont
  {Stadler}},\ }\href@noop {} {\bibfield  {journal} {\bibinfo  {journal} {Phys.
  Rev. B}\ }\textbf {\bibinfo {volume} {88}},\ \bibinfo {pages} {035418}
  (\bibinfo {year} {2013})}\BibitemShut {NoStop}%
\bibitem [{\citenamefont {B\^aldea}(2012)}]{baldea_transition_2012}%
  \BibitemOpen
  \bibfield  {author} {\bibinfo {author} {\bibfnamefont {I.}~\bibnamefont
  {B\^aldea}},\ }\href@noop {} {\bibfield  {journal} {\bibinfo  {journal} {EPL
  (Europhysics Letters)}\ }\textbf {\bibinfo {volume} {98}},\ \bibinfo {pages}
  {17010} (\bibinfo {year} {2012})}\BibitemShut {NoStop}%
\bibitem [{\citenamefont {Mera}\ and\ \citenamefont
  {Niquet}(2010)}]{mera_are_2010}%
  \BibitemOpen
  \bibfield  {author} {\bibinfo {author} {\bibfnamefont {H.}~\bibnamefont
  {Mera}}\ and\ \bibinfo {author} {\bibfnamefont {Y.~M.}\ \bibnamefont
  {Niquet}},\ }\href@noop {} {\bibfield  {journal} {\bibinfo  {journal} {Phys.
  Rev. Lett.}\ }\textbf {\bibinfo {volume} {105}},\ \bibinfo {pages} {216408}
  (\bibinfo {year} {2010})}\BibitemShut {NoStop}%
\bibitem [{\citenamefont {Mera}\ \emph {et~al.}(2010)\citenamefont {Mera},
  \citenamefont {Kaasbjerg}, \citenamefont {Niquet},\ and\ \citenamefont
  {Stefanucci}}]{mera_assessing_2010}%
  \BibitemOpen
  \bibfield  {author} {\bibinfo {author} {\bibfnamefont {H.}~\bibnamefont
  {Mera}}, \bibinfo {author} {\bibfnamefont {K.}~\bibnamefont {Kaasbjerg}},
  \bibinfo {author} {\bibfnamefont {Y.~M.}\ \bibnamefont {Niquet}}, \ and\
  \bibinfo {author} {\bibfnamefont {G.}~\bibnamefont {Stefanucci}},\
  }\href@noop {} {\bibfield  {journal} {\bibinfo  {journal} {Phys. Rev. B}\
  }\textbf {\bibinfo {volume} {81}},\ \bibinfo {pages} {035110} (\bibinfo
  {year} {2010})}\BibitemShut {NoStop}%
\bibitem [{\citenamefont
  {B\^aldea}(2014{\natexlab{a}})}]{baldea_quantifying_2014}%
  \BibitemOpen
  \bibfield  {author} {\bibinfo {author} {\bibfnamefont {I.}~\bibnamefont
  {B\^aldea}},\ }\href@noop {} {\bibfield  {journal} {\bibinfo  {journal}
  {Nanotechnology}\ }\textbf {\bibinfo {volume} {25}},\ \bibinfo {pages}
  {455202} (\bibinfo {year} {2014}{\natexlab{a}})}\BibitemShut {NoStop}%
\bibitem [{\citenamefont {B\^aldea}(2014{\natexlab{b}})}]{baldea_quantum_2014}%
  \BibitemOpen
  \bibfield  {author} {\bibinfo {author} {\bibfnamefont {I.}~\bibnamefont
  {B\^aldea}},\ }\href@noop {} {\bibfield  {journal} {\bibinfo  {journal}
  {Faraday Discussions}\ }\textbf {\bibinfo {volume} {174}},\ \bibinfo {pages}
  {37} (\bibinfo {year} {2014}{\natexlab{b}})}\BibitemShut {NoStop}%
\bibitem [{\citenamefont {Shi}\ \emph {et~al.}(2006)\citenamefont {Shi},
  \citenamefont {Dai}, \citenamefont {Zheng},\ and\ \citenamefont
  {Zeng}}]{shi_ab_2006}%
  \BibitemOpen
  \bibfield  {author} {\bibinfo {author} {\bibfnamefont {X.~Q.}\ \bibnamefont
  {Shi}}, \bibinfo {author} {\bibfnamefont {Z.~X.}\ \bibnamefont {Dai}},
  \bibinfo {author} {\bibfnamefont {X.~H.}\ \bibnamefont {Zheng}}, \ and\
  \bibinfo {author} {\bibfnamefont {Z.}~\bibnamefont {Zeng}},\ }\href@noop {}
  {\bibfield  {journal} {\bibinfo  {journal} {J. Phys. Chem. B}\ }\textbf
  {\bibinfo {volume} {110}},\ \bibinfo {pages} {16902} (\bibinfo {year}
  {2006})}\BibitemShut {NoStop}%
\bibitem [{\citenamefont {Gao}\ \emph {et~al.}(2013)\citenamefont {Gao},
  \citenamefont {Li},\ and\ \citenamefont {Jiang}}]{gao_comparable_2013}%
  \BibitemOpen
  \bibfield  {author} {\bibinfo {author} {\bibfnamefont {N.}~\bibnamefont
  {Gao}}, \bibinfo {author} {\bibfnamefont {J.~C.}\ \bibnamefont {Li}}, \ and\
  \bibinfo {author} {\bibfnamefont {Q.}~\bibnamefont {Jiang}},\ }\href@noop {}
  {\bibfield  {journal} {\bibinfo  {journal} {Appl. Phys. Lett.}\ }\textbf
  {\bibinfo {volume} {103}},\ \bibinfo {pages} {263108} (\bibinfo {year}
  {2013})}\BibitemShut {NoStop}%
\bibitem [{\citenamefont {Guti\'errez}\ \emph {et~al.}(2003)\citenamefont
  {Guti\'errez}, \citenamefont {Grossmann},\ and\ \citenamefont
  {Schmidt}}]{gutierrez_conductance_2003}%
  \BibitemOpen
  \bibfield  {author} {\bibinfo {author} {\bibfnamefont {R.}~\bibnamefont
  {Guti\'errez}}, \bibinfo {author} {\bibfnamefont {F.}~\bibnamefont
  {Grossmann}}, \ and\ \bibinfo {author} {\bibfnamefont {R.}~\bibnamefont
  {Schmidt}},\ }\href@noop {} {\bibfield  {journal} {\bibinfo  {journal}
  {{ChemPhysChem}}\ }\textbf {\bibinfo {volume} {4}},\ \bibinfo {pages} {1252}
  (\bibinfo {year} {2003})}\BibitemShut {NoStop}%
\bibitem [{\citenamefont {Sen}\ \emph {et~al.}(2013)\citenamefont {Sen},
  \citenamefont {Lin},\ and\ \citenamefont {Kaun}}]{sen_single-molecule_2013}%
  \BibitemOpen
  \bibfield  {author} {\bibinfo {author} {\bibfnamefont {A.}~\bibnamefont
  {Sen}}, \bibinfo {author} {\bibfnamefont {C.-J.}\ \bibnamefont {Lin}}, \ and\
  \bibinfo {author} {\bibfnamefont {C.-C.}\ \bibnamefont {Kaun}},\ }\href@noop
  {} {\bibfield  {journal} {\bibinfo  {journal} {J. Phys. Chem. C}\ }\textbf
  {\bibinfo {volume} {117}},\ \bibinfo {pages} {13676} (\bibinfo {year}
  {2013})}\BibitemShut {NoStop}%
\bibitem [{\citenamefont {Caliskan}\ and\ \citenamefont
  {Laref}(2014)}]{caliskan_spin_2014}%
  \BibitemOpen
  \bibfield  {author} {\bibinfo {author} {\bibfnamefont {S.}~\bibnamefont
  {Caliskan}}\ and\ \bibinfo {author} {\bibfnamefont {A.}~\bibnamefont
  {Laref}},\ }\href@noop {} {\bibfield  {journal} {\bibinfo  {journal} {Sci.
  Rep.}\ }\textbf {\bibinfo {volume} {4}} (\bibinfo {year} {2014})}\BibitemShut
  {NoStop}%
\bibitem [{\citenamefont {Kaliginedi}\ \emph {et~al.}(2012)\citenamefont
  {Kaliginedi}, \citenamefont {Moreno-Garc\'ia}, \citenamefont {Valkenier},
  \citenamefont {Hong}, \citenamefont {Garc\'ia-Su\'arez}, \citenamefont
  {Buiter}, \citenamefont {Otten}, \citenamefont {Hummelen}, \citenamefont
  {Lambert},\ and\ \citenamefont {Wandlowski}}]{kaliginedi_correlations_2012}%
  \BibitemOpen
  \bibfield  {author} {\bibinfo {author} {\bibfnamefont {V.}~\bibnamefont
  {Kaliginedi}}, \bibinfo {author} {\bibfnamefont {P.}~\bibnamefont
  {Moreno-Garc\'ia}}, \bibinfo {author} {\bibfnamefont {H.}~\bibnamefont
  {Valkenier}}, \bibinfo {author} {\bibfnamefont {W.}~\bibnamefont {Hong}},
  \bibinfo {author} {\bibfnamefont {V.~M.}\ \bibnamefont {Garc\'ia-Su\'arez}},
  \bibinfo {author} {\bibfnamefont {P.}~\bibnamefont {Buiter}}, \bibinfo
  {author} {\bibfnamefont {J.~L.~H.}\ \bibnamefont {Otten}}, \bibinfo {author}
  {\bibfnamefont {J.~C.}\ \bibnamefont {Hummelen}}, \bibinfo {author}
  {\bibfnamefont {C.~J.}\ \bibnamefont {Lambert}}, \ and\ \bibinfo {author}
  {\bibfnamefont {T.}~\bibnamefont {Wandlowski}},\ }\href@noop {} {\bibfield
  {journal} {\bibinfo  {journal} {J. Am. Chem. Soc.}\ }\textbf {\bibinfo
  {volume} {134}},\ \bibinfo {pages} {5262} (\bibinfo {year}
  {2012})}\BibitemShut {NoStop}%
\bibitem [{\citenamefont {Perrin}\ \emph {et~al.}(2014)\citenamefont {Perrin},
  \citenamefont {Frisenda}, \citenamefont {Koole}, \citenamefont {Seldenthuis},
  \citenamefont {Gil}, \citenamefont {Valkenier}, \citenamefont {Hummelen},
  \citenamefont {Renaud}, \citenamefont {Grozema}, \citenamefont {Thijssen},
  \citenamefont {Duli\'c},\ and\ \citenamefont {van~der
  Zant}}]{perrin_large_2014}%
  \BibitemOpen
  \bibfield  {author} {\bibinfo {author} {\bibfnamefont {M.~L.}\ \bibnamefont
  {Perrin}}, \bibinfo {author} {\bibfnamefont {R.}~\bibnamefont {Frisenda}},
  \bibinfo {author} {\bibfnamefont {M.}~\bibnamefont {Koole}}, \bibinfo
  {author} {\bibfnamefont {J.~S.}\ \bibnamefont {Seldenthuis}}, \bibinfo
  {author} {\bibfnamefont {J.~A.~C.}\ \bibnamefont {Gil}}, \bibinfo {author}
  {\bibfnamefont {H.}~\bibnamefont {Valkenier}}, \bibinfo {author}
  {\bibfnamefont {J.~C.}\ \bibnamefont {Hummelen}}, \bibinfo {author}
  {\bibfnamefont {N.}~\bibnamefont {Renaud}}, \bibinfo {author} {\bibfnamefont
  {F.~C.}\ \bibnamefont {Grozema}}, \bibinfo {author} {\bibfnamefont {J.~M.}\
  \bibnamefont {Thijssen}}, \bibinfo {author} {\bibfnamefont {D.}~\bibnamefont
  {Duli\'c}}, \ and\ \bibinfo {author} {\bibfnamefont {H.~S.~J.}\ \bibnamefont
  {van~der Zant}},\ }\href@noop {} {\bibfield  {journal} {\bibinfo  {journal}
  {Nat. Nanotechnol.}\ }\textbf {\bibinfo {volume} {9}},\ \bibinfo {pages}
  {830} (\bibinfo {year} {2014})}\BibitemShut {NoStop}%
\bibitem [{\citenamefont {Strange}\ and\ \citenamefont
  {Thygesen}(2011)}]{strange_towards_2011}%
  \BibitemOpen
  \bibfield  {author} {\bibinfo {author} {\bibfnamefont {M.}~\bibnamefont
  {Strange}}\ and\ \bibinfo {author} {\bibfnamefont {K.~S.}\ \bibnamefont
  {Thygesen}},\ }\href@noop {} {\bibfield  {journal} {\bibinfo  {journal}
  {Beilstein J. Nanotechnol.}\ }\textbf {\bibinfo {volume} {2}},\ \bibinfo
  {pages} {746} (\bibinfo {year} {2011})}\BibitemShut {NoStop}%
\bibitem [{\citenamefont {Ning}\ \emph {et~al.}(2007)\citenamefont {Ning},
  \citenamefont {Li}, \citenamefont {Shen}, \citenamefont {Qian}, \citenamefont
  {Hou}, \citenamefont {Rocha},\ and\ \citenamefont
  {Sanvito}}]{ning_first-principles_2007}%
  \BibitemOpen
  \bibfield  {author} {\bibinfo {author} {\bibfnamefont {J.}~\bibnamefont
  {Ning}}, \bibinfo {author} {\bibfnamefont {R.}~\bibnamefont {Li}}, \bibinfo
  {author} {\bibfnamefont {X.}~\bibnamefont {Shen}}, \bibinfo {author}
  {\bibfnamefont {Z.}~\bibnamefont {Qian}}, \bibinfo {author} {\bibfnamefont
  {S.}~\bibnamefont {Hou}}, \bibinfo {author} {\bibfnamefont {A.~R.}\
  \bibnamefont {Rocha}}, \ and\ \bibinfo {author} {\bibfnamefont
  {S.}~\bibnamefont {Sanvito}},\ }\href@noop {} {\bibfield  {journal} {\bibinfo
   {journal} {Nanotechnology}\ }\textbf {\bibinfo {volume} {18}},\ \bibinfo
  {pages} {345203} (\bibinfo {year} {2007})}\BibitemShut {NoStop}%
\bibitem [{\citenamefont {Marzari}\ and\ \citenamefont
  {Vanderbilt}(1997)}]{marzari_maximally_1997}%
  \BibitemOpen
  \bibfield  {author} {\bibinfo {author} {\bibfnamefont {N.}~\bibnamefont
  {Marzari}}\ and\ \bibinfo {author} {\bibfnamefont {D.}~\bibnamefont
  {Vanderbilt}},\ }\href@noop {} {\bibfield  {journal} {\bibinfo  {journal}
  {Phys. Rev. B}\ }\textbf {\bibinfo {volume} {56}},\ \bibinfo {pages} {12847}
  (\bibinfo {year} {1997})}\BibitemShut {NoStop}%
\bibitem [{\citenamefont {Perdew}\ \emph {et~al.}(1996)\citenamefont {Perdew},
  \citenamefont {Burke},\ and\ \citenamefont {Ernzerhof}}]{PBE}%
  \BibitemOpen
  \bibfield  {author} {\bibinfo {author} {\bibfnamefont {J.~P.}\ \bibnamefont
  {Perdew}}, \bibinfo {author} {\bibfnamefont {K.}~\bibnamefont {Burke}}, \
  and\ \bibinfo {author} {\bibfnamefont {M.}~\bibnamefont {Ernzerhof}},\
  }\href@noop {} {\bibfield  {journal} {\bibinfo  {journal} {Phys. Rev. Lett.}\
  }\textbf {\bibinfo {volume} {77}},\ \bibinfo {pages} {3865} (\bibinfo {year}
  {1996})}\BibitemShut {NoStop}%
\bibitem [{\citenamefont {Gonze}\ \emph {et~al.}(2009)\citenamefont {Gonze},
  \citenamefont {Amadon}, \citenamefont {Anglade}, \citenamefont {Beuken},
  \citenamefont {Bottin}, \citenamefont {Boulanger}, \citenamefont {Bruneval},
  \citenamefont {Caliste}, \citenamefont {Caracas}, \citenamefont {C\^ot\'e},
  \citenamefont {Deutsch}, \citenamefont {Genovese}, \citenamefont {Ghosez},
  \citenamefont {Giantomassi}, \citenamefont {Goedecker}, \citenamefont
  {Hamann}, \citenamefont {Hermet}, \citenamefont {Jollet}, \citenamefont
  {Jomard}, \citenamefont {Leroux}, \citenamefont {Mancini}, \citenamefont
  {Mazevet}, \citenamefont {Oliveira}, \citenamefont {Onida}, \citenamefont
  {Pouillon}, \citenamefont {Rangel}, \citenamefont {Rignanese}, \citenamefont
  {Sangalli}, \citenamefont {Shaltaf}, \citenamefont {Torrent}, \citenamefont
  {Verstraete}, \citenamefont {Zerah},\ and\ \citenamefont
  {Zwanziger}}]{abinit}%
  \BibitemOpen
  \bibfield  {author} {\bibinfo {author} {\bibfnamefont {X.}~\bibnamefont
  {Gonze}}, \bibinfo {author} {\bibfnamefont {B.}~\bibnamefont {Amadon}},
  \bibinfo {author} {\bibfnamefont {P.~M.}\ \bibnamefont {Anglade}}, \bibinfo
  {author} {\bibfnamefont {J.~M.}\ \bibnamefont {Beuken}}, \bibinfo {author}
  {\bibfnamefont {F.}~\bibnamefont {Bottin}}, \bibinfo {author} {\bibfnamefont
  {P.}~\bibnamefont {Boulanger}}, \bibinfo {author} {\bibfnamefont
  {F.}~\bibnamefont {Bruneval}}, \bibinfo {author} {\bibfnamefont
  {D.}~\bibnamefont {Caliste}}, \bibinfo {author} {\bibfnamefont
  {R.}~\bibnamefont {Caracas}}, \bibinfo {author} {\bibfnamefont
  {M.}~\bibnamefont {C\^ot\'e}}, \bibinfo {author} {\bibfnamefont
  {T.}~\bibnamefont {Deutsch}}, \bibinfo {author} {\bibfnamefont
  {L.}~\bibnamefont {Genovese}}, \bibinfo {author} {\bibfnamefont
  {P.}~\bibnamefont {Ghosez}}, \bibinfo {author} {\bibfnamefont
  {M.}~\bibnamefont {Giantomassi}}, \bibinfo {author} {\bibfnamefont
  {S.}~\bibnamefont {Goedecker}}, \bibinfo {author} {\bibfnamefont {D.~R.}\
  \bibnamefont {Hamann}}, \bibinfo {author} {\bibfnamefont {P.}~\bibnamefont
  {Hermet}}, \bibinfo {author} {\bibfnamefont {F.}~\bibnamefont {Jollet}},
  \bibinfo {author} {\bibfnamefont {G.}~\bibnamefont {Jomard}}, \bibinfo
  {author} {\bibfnamefont {S.}~\bibnamefont {Leroux}}, \bibinfo {author}
  {\bibfnamefont {M.}~\bibnamefont {Mancini}}, \bibinfo {author} {\bibfnamefont
  {S.}~\bibnamefont {Mazevet}}, \bibinfo {author} {\bibfnamefont {M.~J.~T.}\
  \bibnamefont {Oliveira}}, \bibinfo {author} {\bibfnamefont {G.}~\bibnamefont
  {Onida}}, \bibinfo {author} {\bibfnamefont {Y.}~\bibnamefont {Pouillon}},
  \bibinfo {author} {\bibfnamefont {T.}~\bibnamefont {Rangel}}, \bibinfo
  {author} {\bibfnamefont {G.~M.}\ \bibnamefont {Rignanese}}, \bibinfo {author}
  {\bibfnamefont {D.}~\bibnamefont {Sangalli}}, \bibinfo {author}
  {\bibfnamefont {R.}~\bibnamefont {Shaltaf}}, \bibinfo {author} {\bibfnamefont
  {M.}~\bibnamefont {Torrent}}, \bibinfo {author} {\bibfnamefont {M.~J.}\
  \bibnamefont {Verstraete}}, \bibinfo {author} {\bibfnamefont
  {G.}~\bibnamefont {Zerah}}, \ and\ \bibinfo {author} {\bibfnamefont {J.~W.}\
  \bibnamefont {Zwanziger}},\ }\href@noop {} {\bibfield  {journal} {\bibinfo
  {journal} {Comput. Phys. Commun.}\ }\textbf {\bibinfo {volume} {180}},\
  \bibinfo {pages} {2582} (\bibinfo {year} {2009})}\BibitemShut {NoStop}%
\bibitem [{wan()}]{want}%
  \BibitemOpen
  \href {http://www.wannier-transport.org} {\enquote {\bibinfo {title} {{WanT}
  code by {A. Ferretti, L. Agapito, A. Calzolari, and M. Buongiorno
  Nardelli}.}}\ }\BibitemShut {NoStop}%
\bibitem [{\citenamefont {Calzolari}\ \emph {et~al.}(2004)\citenamefont
  {Calzolari}, \citenamefont {Marzari}, \citenamefont {Souza},\ and\
  \citenamefont {Buongiorno~Nardelli}}]{calzolari_want_2004}%
  \BibitemOpen
  \bibfield  {author} {\bibinfo {author} {\bibfnamefont {A.}~\bibnamefont
  {Calzolari}}, \bibinfo {author} {\bibfnamefont {N.}~\bibnamefont {Marzari}},
  \bibinfo {author} {\bibfnamefont {I.}~\bibnamefont {Souza}}, \ and\ \bibinfo
  {author} {\bibfnamefont {M.}~\bibnamefont {Buongiorno~Nardelli}},\
  }\href@noop {} {\bibfield  {journal} {\bibinfo  {journal} {Phys. Rev. B}\
  }\textbf {\bibinfo {volume} {69}},\ \bibinfo {pages} {035108} (\bibinfo
  {year} {2004})}\BibitemShut {NoStop}%
\bibitem [{\citenamefont {M\"uller}(2006)}]{muller_effect_2006}%
  \BibitemOpen
  \bibfield  {author} {\bibinfo {author} {\bibfnamefont {K.-H.}\ \bibnamefont
  {M\"uller}},\ }\href@noop {} {\bibfield  {journal} {\bibinfo  {journal}
  {Phys. Rev. B}\ }\textbf {\bibinfo {volume} {73}},\ \bibinfo {pages} {045403}
  (\bibinfo {year} {2006})}\BibitemShut {NoStop}%
\bibitem [{\citenamefont {French}\ \emph {et~al.}(2013)\citenamefont {French},
  \citenamefont {Iacovella}, \citenamefont {Rungger}, \citenamefont {Souza},
  \citenamefont {Sanvito},\ and\ \citenamefont
  {Cummings}}]{french_structural_2013}%
  \BibitemOpen
  \bibfield  {author} {\bibinfo {author} {\bibfnamefont {W.~R.}\ \bibnamefont
  {French}}, \bibinfo {author} {\bibfnamefont {C.~R.}\ \bibnamefont
  {Iacovella}}, \bibinfo {author} {\bibfnamefont {I.}~\bibnamefont {Rungger}},
  \bibinfo {author} {\bibfnamefont {A.~M.}\ \bibnamefont {Souza}}, \bibinfo
  {author} {\bibfnamefont {S.}~\bibnamefont {Sanvito}}, \ and\ \bibinfo
  {author} {\bibfnamefont {P.~T.}\ \bibnamefont {Cummings}},\ }\href@noop {}
  {\bibfield  {journal} {\bibinfo  {journal} {J. Phys. Chem. Lett.}\ }\textbf
  {\bibinfo {volume} {4}},\ \bibinfo {pages} {887} (\bibinfo {year}
  {2013})}\BibitemShut {NoStop}%
\bibitem [{\citenamefont {French}\ \emph {et~al.}(2012)\citenamefont {French},
  \citenamefont {Iacovella},\ and\ \citenamefont
  {Cummings}}]{french_large-scale_2012}%
  \BibitemOpen
  \bibfield  {author} {\bibinfo {author} {\bibfnamefont {W.~R.}\ \bibnamefont
  {French}}, \bibinfo {author} {\bibfnamefont {C.~R.}\ \bibnamefont
  {Iacovella}}, \ and\ \bibinfo {author} {\bibfnamefont {P.~T.}\ \bibnamefont
  {Cummings}},\ }\href@noop {} {\bibfield  {journal} {\bibinfo  {journal} {ACS
  Nano}\ }\textbf {\bibinfo {volume} {6}},\ \bibinfo {pages} {2779} (\bibinfo
  {year} {2012})}\BibitemShut {NoStop}%
\bibitem [{\citenamefont {Souza}\ \emph {et~al.}(2014)\citenamefont {Souza},
  \citenamefont {Rungger}, \citenamefont {Pontes}, \citenamefont {Rocha},
  \citenamefont {Silva}, \citenamefont {Schwingenschl\"oegl},\ and\
  \citenamefont {Sanvito}}]{souza_stretching_2014}%
  \BibitemOpen
  \bibfield  {author} {\bibinfo {author} {\bibfnamefont {A.~d.~M.}\
  \bibnamefont {Souza}}, \bibinfo {author} {\bibfnamefont {I.}~\bibnamefont
  {Rungger}}, \bibinfo {author} {\bibfnamefont {R.~B.}\ \bibnamefont {Pontes}},
  \bibinfo {author} {\bibfnamefont {A.~R.}\ \bibnamefont {Rocha}}, \bibinfo
  {author} {\bibfnamefont {A.~J. R.~d.}\ \bibnamefont {Silva}}, \bibinfo
  {author} {\bibfnamefont {U.}~\bibnamefont {Schwingenschl\"oegl}}, \ and\
  \bibinfo {author} {\bibfnamefont {S.}~\bibnamefont {Sanvito}},\ }\href@noop
  {} {\bibfield  {journal} {\bibinfo  {journal} {Nanoscale}\ }\textbf {\bibinfo
  {volume} {6}},\ \bibinfo {pages} {14495} (\bibinfo {year}
  {2014})}\BibitemShut {NoStop}%
\bibitem [{\citenamefont {Nguyen}\ \emph {et~al.}(2014)\citenamefont {Nguyen},
  \citenamefont {Szyja},\ and\ \citenamefont
  {Doltsinis}}]{nguyen_electric_2014}%
  \BibitemOpen
  \bibfield  {author} {\bibinfo {author} {\bibfnamefont {H.~C.}\ \bibnamefont
  {Nguyen}}, \bibinfo {author} {\bibfnamefont {B.~M.}\ \bibnamefont {Szyja}}, \
  and\ \bibinfo {author} {\bibfnamefont {N.~L.}\ \bibnamefont {Doltsinis}},\
  }\href@noop {} {\bibfield  {journal} {\bibinfo  {journal} {Phys. Rev. B}\
  }\textbf {\bibinfo {volume} {90}},\ \bibinfo {pages} {115440} (\bibinfo
  {year} {2014})}\BibitemShut {NoStop}%
\bibitem [{\citenamefont {Qian}\ \emph {et~al.}(2010)\citenamefont {Qian},
  \citenamefont {Li},\ and\ \citenamefont {Yip}}]{qian_calculating_2010}%
  \BibitemOpen
  \bibfield  {author} {\bibinfo {author} {\bibfnamefont {X.}~\bibnamefont
  {Qian}}, \bibinfo {author} {\bibfnamefont {J.}~\bibnamefont {Li}}, \ and\
  \bibinfo {author} {\bibfnamefont {S.}~\bibnamefont {Yip}},\ }\href@noop {}
  {\bibfield  {journal} {\bibinfo  {journal} {Phys. Rev. B}\ }\textbf {\bibinfo
  {volume} {82}},\ \bibinfo {pages} {195442} (\bibinfo {year}
  {2010})}\BibitemShut {NoStop}%
\bibitem [{\citenamefont {Sakuma}(2013)}]{sakuma_symmetry-adapted_2013}%
  \BibitemOpen
  \bibfield  {author} {\bibinfo {author} {\bibfnamefont {R.}~\bibnamefont
  {Sakuma}},\ }\href@noop {} {\bibfield  {journal} {\bibinfo  {journal} {Phys.
  Rev. B}\ }\textbf {\bibinfo {volume} {87}},\ \bibinfo {pages} {235109}
  (\bibinfo {year} {2013})}\BibitemShut {NoStop}%
\bibitem [{\citenamefont {Agapito}\ \emph {et~al.}(2013)\citenamefont
  {Agapito}, \citenamefont {Ferretti}, \citenamefont {Calzolari}, \citenamefont
  {Curtarolo},\ and\ \citenamefont
  {Buongiorno~Nardelli}}]{agapito_effective_2013}%
  \BibitemOpen
  \bibfield  {author} {\bibinfo {author} {\bibfnamefont {L.~A.}\ \bibnamefont
  {Agapito}}, \bibinfo {author} {\bibfnamefont {A.}~\bibnamefont {Ferretti}},
  \bibinfo {author} {\bibfnamefont {A.~o.}\ \bibnamefont {Calzolari}}, \bibinfo
  {author} {\bibfnamefont {S.}~\bibnamefont {Curtarolo}}, \ and\ \bibinfo
  {author} {\bibfnamefont {M.}~\bibnamefont {Buongiorno~Nardelli}},\
  }\href@noop {} {\bibfield  {journal} {\bibinfo  {journal} {Phys. Rev. B}\
  }\textbf {\bibinfo {volume} {88}},\ \bibinfo {pages} {165127} (\bibinfo
  {year} {2013})}\BibitemShut {NoStop}%
\bibitem [{\citenamefont {Lherbier}\ \emph {et~al.}(2013)\citenamefont
  {Lherbier}, \citenamefont {Roche}, \citenamefont {Restrepo}, \citenamefont
  {Niquet}, \citenamefont {Delcorte},\ and\ \citenamefont
  {Charlier}}]{lherbier_highly_2013}%
  \BibitemOpen
  \bibfield  {author} {\bibinfo {author} {\bibfnamefont {A.}~\bibnamefont
  {Lherbier}}, \bibinfo {author} {\bibfnamefont {S.}~\bibnamefont {Roche}},
  \bibinfo {author} {\bibfnamefont {O.~A.}\ \bibnamefont {Restrepo}}, \bibinfo
  {author} {\bibfnamefont {Y.-M.}\ \bibnamefont {Niquet}}, \bibinfo {author}
  {\bibfnamefont {A.}~\bibnamefont {Delcorte}}, \ and\ \bibinfo {author}
  {\bibfnamefont {J.-C.}\ \bibnamefont {Charlier}},\ }\href@noop {} {\bibfield
  {journal} {\bibinfo  {journal} {Nano Res.}\ }\textbf {\bibinfo {volume}
  {6}},\ \bibinfo {pages} {326} (\bibinfo {year} {2013})}\BibitemShut {NoStop}%
\end{thebibliography}%

\end{document}